\documentclass[prd,floats,preprint,superscriptaddress,eqsecnum,floatfix,nofootinbib,12pt,tightenlines]{revtex4}
\usepackage{amsmath}
\usepackage{graphics, graphicx}
\usepackage{epsfig}
\usepackage{ulem}
\usepackage{color}

\def\bu{BU\ }
\def\td{TD\ }
\def\1p{{(1p)}}
\def\be{\begin{equation}}
\def\ee{\end{equation}}
\def\beq{\begin{eqnarray}}
\def\eeq{\end{eqnarray}}

\def\p0{\phi_0}
\def\z0{\zeta_0}
\def\zs{\zeta_s(z)}
\def\aoi{D^{\ge 1}}
\def\aoh{D_h^{\ge 1}}

\def\epa{{(0)}}
\def\epb{{(2)}}

\def\vf{}

\begin{document}

\title{The No-Boundary Measure in the Regime of Eternal Inflation}

\author{James  Hartle}
\affiliation{Department of Physics, University of California, Santa Barbara,  93106, USA}
\author{S.W. Hawking}
\affiliation{DAMTP, CMS, Wilberforce Road, CB3 0WA Cambridge, UK}
\author{Thomas Hertog}
\affiliation{APC, UMR 7164 (CNRS, Universit\'e Paris 7), 10 rue A.Domon et L.Duquet, 75205 Paris, France\\ {\it and}\\
International Solvay Institutes, Boulevard du Triomphe, ULB -- C.P. 231, 1050 Brussels, Belgium}

\bibliographystyle{unsrt}

\begin{abstract}

The no-boundary wave function (NBWF) specifies a measure for prediction in cosmology that selects inflationary histories and remains well behaved for spatially large or infinite universes. This paper explores the predictions of the NBWF for linear scalar fluctuations about homogeneous and isotropic backgrounds in models with a single scalar field moving in a quadratic potential. We treat both the space-time geometry of the universe and the observers inhabiting it quantum mechanically. We evaluate top-down probabilities for local observations that are conditioned on the NBWF and on part of our data as observers of the universe. For models where the most probable histories  do not have  a regime of eternal inflation, the NBWF predicts homogeneity on large scales, a specific non-Gaussian spectrum of observable fluctuations, and a small amount of inflation in our past. By contrast, for models where the dominant histories have a regime of eternal inflation, the NBWF predicts significant inhomogeneity on scales much larger than the present horizon, a Gaussian spectrum of observable fluctuations, and a long period of inflation in our past. The absence or presence of local non-Gaussianity therefore provides information about the global structure of the universe, assuming the NBWF.

\end{abstract}

\vspace{1cm}

\pacs{98.80.Qc, 98.80.Bp, 98.80.Cq, 04.60.-m [CHECK]}

\maketitle

\section{Introduction}
\label{intro}

Inflation tends to make the universe so large that we can at best observe only a tiny part of it. Even for a closed universe it has been argued that when there is a regime of eternal inflation, inhomogeneities can lead to an infinitely large reheating surface \cite{EIrefs}. Bubble nucleation in false vacuum models,  for instance, leads to inhomogeneous universes that contain infinite open spatial slices of constant density inside the bubbles \cite{CdL80}. The issues that arise for prediction as a consequence of large or infinite sized universes are loosely referred to as the measure problem\footnote{Although it is usually discussed in the context of inflation (see e.g.\cite{measurerefs} for recent work), the measure problem is not specific to inflationary cosmology. Similar issues arise in any theory of cosmology that predicts spatially infinite universes.}. 

This is one of a series of papers  \cite{HHH08a,HHH08b,HH09} devoted to the measure for prediction provided by the no-boundary quantum state of the universe (NBWF) \cite{HH83,Haw84}. If the universe is a quantum mechanical system it has a quantum state. This state predicts probabilities for alternative histories of the universe and everything in it,  including the alternative histories of its spacetime geometry. It seems inevitable that any discussion of prediction in fundamental cosmology should take these probabilities into account. Indeed, it is plausible that, except for an assumption of the typicality of our data,  no measure beyond that supplied by the NBWF is needed for any observational prediction.

Previous papers \cite{HHH08a,HHH08b,HH09} considered the NBWF's predictions in homogeneous, isotropic minisuperspace models. We showed how the no-boundary measure for observations is well defined even in the limit of very large universes provided that the quantum nature of the observer making the observation is taken into account. In this paper we extend this work to consider linear fluctuations in matter and geometry away from homogeneity and isotropy. This enables us to consider predictions for observable quantities such as those connected with the cosmic microwave background (CMB). 
We continue to model the matter by a single scalar field moving in a quadratic potential. We also continue with our simple model of an observer as a physical system characterized by data $D$ and a probability $p_E(D)$ to exist in any one Hubble volume on spacelike surfaces specified by $D$. 

We will describe our assumptions and procedures for calculation in Section \ref{observations}. But one crucial distinction should be mentioned at the outset --- the difference between top-down (\td) and bottom-up (\bu) probabilities \cite{Hawking06}.  

By itself, the NBWF predicts the probabilities for the alternative, four-dimensional, classical histories that the universe may exhibit. We call these the \bu probabilities for the classical ensemble. However, we do not observe entire histories. Instead our observations are restricted to a light cone located somewhere in the universe and extending over roughly a Hubble volume. 
Predictions for observations in cosmology are necessarily conditioned on a description of the local observational situation in addition to the NBWF. For instance an observation of the CMB spectrum depends on {\it when} and {\it where} the observation is made in the history of the universe. In general, probabilities for {\it our} observations are conditioned on some part of our data $D$ and predict other properties of the universe. Probabilities conditioned on some part of our data are called \td probabilities. They can differ significantly from the \bu probabilities for the same alternatives as those for the number of efolds of inflation discussed in \cite{HHH08b}. The probabilities for perturbations that are within our current horizon that will be calculated here provide another illustration of this difference.

After a brief statement of our assumptions and procedures in Section \ref{observations} the paper proceeds to derive the \td probabilities for local observations related to fluctuations as follows: In Sections \ref{classquantfluct} and \ref{bottom-up}  we calculate the classical ensemble of four-dimensional, Lorentzian, homogeneous and isotropic (homo/iso) histories with linear scalar fluctuations predicted by the semiclassical approximation to the  NBWF. The real part of the Euclidean action of the complex saddle-points corresponding to the different histories in the ensemble provides the \bu NBWF probabilities of both the homo/iso backgrounds and their perturbations viewed as global features of the universe. From these bottom-up probabilities one can obtain predictions for local observations such as the CMB temperature anisotropies. This is done in Sections \ref{top-down} where we calculate the top-down probabilities for observing different perturbations inside our current horizon. We find a slightly non-Gaussian spectrum of perturbations on currently observable scales in models where the most probable histories  do not have  a regime of eternal inflation. By contrast, for models where the dominant histories have a regime of eternal inflation, we find the NBWF predicts a Gaussian spectrum of observable fluctuations. In Section \ref{topdown_inei} we comment on backreaction effects in the regime of eternal inflation, and argue that these are unlikely to change the above results. Finally in Section \ref{conclu} we present our conclusions.

\section{From the NBWF to Probabilities for Our Observations}
\label{observations}\label{sec2}

This section sets out our assumptions and procedures for calculating the probabilities for our observations from a quantum state of the universe. These are then illustrated in a simple model.

\subsection{Framework}
{\it A quantum universe with a quantum state.} We assume that the universe is a closed quantum mechanical system with a particular quantum state. 
That state is taken to be the NBWF. The state predicts probabilities for the individual members of  decoherent sets of alternative, coarse-grained, four-dimensional histories of the universe and its contents according to the principles of generalized quantum theory \cite{Hartle95}. We call these bottom-up (BU) probabilities.  

{\it Bottom-up probabilities for the classical ensemble.} In particular, the state predicts the probabilities for the classical ensemble consisting of four-dimensional  alternative histories with high probabilities for correlations in time governed by classical equations of motion.  In \cite{HHH08a,HHH08b,HHrules} we described how to calculate a semiclassical approximation for the probabilities of these histories from the semiclassical approximation to the NBWF assuming decoherence in  an appropriate coarse graining. In this paper we restrict attention to probabilities of alternatives that can be  defined in terms of these classical histories. Simple examples are the probabilities for the number of inflationary efolds or for the size of  fluctuations away from homogeneity and isotropy. 

{\it Top-down probabilities for observations.}
 Probabilities for our\footnote{`We', `us', `our' etc refer loosely to the collection of humans engaged in scientific research on cosmology. We will not need a more precise definition.} observations are not probabilities for a four-dimensional history of the universe. Rather they are probabilities for local alternatives at a particular time and place in a classical history.
They are constructed from the \bu probabilities supplied by the NBWF by conditioning on at least that part of our data that describes what we know of our location in spacetime. We call  probabilities conditioned on all or part of our data top-down (TD) probabilities. 

{\it Observers as quantum systems.} As observers we are quantum systems within the universe characterized at an appropriate coarse-grained level by the data $D$ that we possess --- including a physical description of ourselves. We arose from physical processes that occurred over the universe's history. We are therefore not certain to exist in the universe, Indeed, there is only a very tiny probability $p_E(D)$ for an instance of the data $D$ in any Hubble volume. However, in a very large universe the probability becomes significant that the data $D$ are replicated elsewhere.  All we know for sure about the universe is that it exhibits at least one instance of the  data $D$ --- a situation we abbreviate by $\aoi$.  \td probabilities  can differ significantly from \bu probabilities  for the same alternatives when these facts about observers are taken into account as we now illustrate in a very simple model.

\subsection{Procedures Illustrated by a Simple Model}

Consider a toy model universe consisting of a number of boxes --- `Hubble volumes'. We consider these at a single moment of time. There are $K$ physical degrees of freedom $z_1, \cdots z_K$ each constrained to be the same in all Hubble volumes.  We denote them collectively by $z\equiv (z_1, \cdots , z_K)$. The quantum state supplies \bu probabilities\footnote{Here and throughout we do not distinguish notationally between probabilities and probability densities.} $p(z)$ for the values of the $z_i$. The fields $z_i$ are crudely analogous to the fluctuations away from homogeneity and isotropy that we will consider later. The number of Hubble volumes $N_h$ depends on $z$, $N_h(z)$, as it would for a fluctuation in geometry. Observers in the Hubble volumes can measure $z$. The probability that there is an observer with data $D$ in any Hubble volume is $p_E(D)$.  With this simple model we will be able to illustrate many of our procedures and results without getting bogged down in the elegant but complex technology of cosmological perturbation theory.  

We distinguish between local and global predictions. Local predictions are for features of the universe inside a Hubble volume --- features that we could in principle observe in ours.  Examples are the CMB correlation functions. Global predictions are for features of our universe that may extend outside our Hubble volume or beyond the present time. Examples are predictions of the number of efolds of inflation in the past or  inhomogeneities outside the present horizon. 

Probabilities for the results of our observations are for local features of the universe conditioned on what we know about it --- our data $D$. All we know for certain from our data is that the universe exhibits at least one instance of it, $\aoi$. In the present model we can consider the probabilities for the value of $z$ in a Hubble  volume given $\aoi$. But we are not just interested in these probabilities in any Hubble volume; we are interested in them in {\it our} Hubble volume. We discuss how to handle such first person questions generally in Section \ref{thirdtofirst}. But in cases where there is a symmetry between Hubble volumes there is a shortcut to the answer.

This simple model has  such a symmetry --- all the Hubble volumes are the same.  The probability for a value of $z$ in our Hubble volume given $\aoi$  is therefore the same as the probability that the universe exhibits a value of $z$ in any Hubble volume given $\aoi$.  And since $z$ is constant over the universe that is the same as the probability $p(z|\aoi)$ that the universe has a value of $z$ given $\aoi$. 

This can  can be efficiently computed by starting from the relation
\begin{equation}
\label{reverse}
p(z|\aoi) = \frac{p(z,\aoi)}{p(\aoi)} = \frac{p(\aoi|z)p(z)}{p(\aoi)}  .
\end{equation}
The probability $p(\aoi|z)$ that there is at least one instance of $D$ in the universe given $z$ is $1$ minus the probability that there are no instances in any Hubble volume. This is 
\begin{equation}
\label{atleast1}
p(\aoi|z) =1-(1-p_E(D))^{N_h(z)} .
\end{equation}
Combining \eqref{reverse} and \eqref{atleast1} gives
\begin{equation}
\label{thirdperson}
p(z|\aoi) = \frac{[1-(1-p_E(D))^{N_h(z)} ] p(z)}{\int dz [1-(1-p_E(D))^{N_h(z)} ] p(z)} .
\end{equation}
This is the probability for our observations of $z$ in this very simple model. 

\subsection{Gaussianity and Non-Gaussianity}
Eq \eqref{thirdperson} for the probability of our observations of $z$ simplifies in two important limits. First, it simplifies  when 
$p_E(D)N_h(z) \ll 1$ for the whole range of $z$, that is, in the limit in which we are rare in the universe. Then we have
\begin{equation}
\label{rare}
p(z|\aoi) \approx \frac{N_h(z)p(z)}{\int dz N_h(z)p(z)} .
\end{equation}
The difficult to estimate\footnote{For a discussion of ways to bound $p_E(D)$ see \cite{HH09}.}  probability $p_E(D)$ has cancelled out. $N_h(z)$ would also cancel were it independent of $z$ leaving the probability for observation of a value  $z$ equal to the \bu probability that the universe has that value. 

But if  $N_h(z)$ depends on  $z$ the probabilities for observation will  differ from the bottom-up probabilities. In particular suppose the bottom-up probabilities $p(z) $ are Gaussian, that is a product of terms of the form $ \exp(-{\rm const} \ z_i^2)$.  Then the probabilities for observing $z$ will not be Gaussian. The \td probabilities for values of $z$ that make the universe larger are enhanced over their \bu values because in a larger universe there are more places for our data $D$ to be. 

The second limit in which $p(z|\aoi)$ is independent of $p_E(D)$ is when $p_E(D)N_h(z) \gg 1$ for the whole range of $z$. This is the limit where our universe is so large that our data are common. Then,
\begin{equation}
\label{common}
p(z|\aoi) \approx p(z) ,
\end{equation}
that is, \td probabilities equal \bu probabilities. As a consequence,  Gaussian \bu probabilities imply a Gaussian distribution for the probabilities of observing $z$.

Thus, from local measurement of the $z_i$ an observer confident of the validity of this simple model could infer something about the size of the universe $N_h$. That does not violate causality. The data $D$ may be assumed to be within our past light cone. But the quantum state predicts non-local correlations between the properties of different Hubble volumes. These can be exploited to make predictions outside our Hubble volume from data inside it. This is not qualitatively different from assuming that the universe is homogeneous and then inferring the mean density outside our Hubble volume from observations inside. 

The common limit \eqref{common} shows that predictions for observations can be defined even when there are an infinite number of Hubble volumes provided that the \bu probabilities are normalized. No `measure' beyond that provided by the quantum state is needed to deal with infinite volumes in these simple models.

\subsection{Detecting Gaussianity}
Suppose that the \bu probabilities for the $z$'s are Gaussian. That is, suppose specifically that,
\begin{equation}
\label{gauss_z}
p(z)  =\prod_i  (2 \pi \sigma^2)^{-1/2}  \exp(- z_i^2/2\sigma^2)  .
\end{equation}
As \eqref{rare} shows, Gaussian \bu probabilities do not necessarily imply Gaussian \td probabilities for observation. But to understand a little more about the tests for {\it non}-Gaussianity let us first consider the common limit \eqref{common} where the \td probabilities are Gaussian.

A complete description of our universe will generally require variables other than those we observe directly. In the absence of observation we may only have probabilities for these variables and the resulting \td probabilities for observation may not have the simple Gaussian form. But, if a Gaussian distribution is predicted for all values of the unknown variables, Gaussian statistics for observation will still be predicted. This elementary but important point can be illustrated with a modest extension of our simple model. 

Suppose that in addition to the $z$'s the widths of the Gaussian distributions in \eqref{gauss_z} depend on a variable $\p0$ so that $\sigma=\sigma(\p0)$. Then the \td probabilities for observation will be given by 
\begin{equation} 
\label{obswp0}
p(z|\aoi) = \int d\p0\ p(z|\aoi, \p0) p(\aoi|\p0) p(\p0)
\end{equation}
where $p(\p0)$ is the \bu probability for the unobserved $\p0$. Even if $p(z|\aoi, \p0)$ is a Gaussian distribution of form \eqref{gauss_z} with $\p0$ dependent $\sigma$'s,  the sum of them in \eqref{obswp0} will not be. Consider  functions of the  $z_i$'s whose expected value vanishes for Gaussian distributions and thus test Gaussianity. An example is the correlation function 
\begin{equation}
\label{3pt}
B_{kj} \equiv \frac{1}{\ell}\sum_i z_i z_{i+k} z_{i+j} . 
\end{equation}
 If the expected value of such functions vanish for each $\p0$ it will also vanish for the sum. The point is that our universe is characterized by some value of $\p0$ even if we have not determined what it is. If Gaussianity is predicted for all values of $\p0$ we predict Gaussianity for our observations of the $z$'s despite our ignorance of $\p0$'s value.

\subsection{From 3rd Person to 1st Person}
\label{thirdtofirst}

The derivation of the probabilities for observation in our simple model relied on a symmetry --- the equivalence of all Hubble volumes. That symmetry allowed us to ignore all the other instances of our data $D$ that a large universe might exhibit and focus on our own. We will rely on an analogous underlying symmetry in our discussion of the fluctuations away from homogeneity and isotropy in Section V. But lest the reader believe that a symmetry is essential to calculating probabilities for observations we present in this section a derivation in the simple model that does not require a symmetry and explicitly takes into account the instances of $D$ beyond our own. 

To begin let us calculate the probability $p(z,n)$ that the universe has the value $z$ and $n$ Hubble volumes with the data $D$. This evidently is
\begin{equation}
p(z,n)=\binom{N_h(z)}{n} \left(p_E(D)\right)^n\left(1-p_E(D)\right)^{N_h(z)-n} p(z) . 
\label{prob_n}
\end{equation}
Since all the Hubble volumes are the same, the sum over locations of the $n$ instances has reduced to the binomial coefficient giving the number of ways of picking $n$ Hubble volumes with observers out of $N_h(z)$ total Hubble volumes. 
The probability $p(z,n)$  is an example of a {\it third person probability} --- a probability for a  feature the universe may exhibit independently of any relation to us.  But we are interested in the {\it first person probability} of what value of $z$ {\it we} will observe. The theory, by itself, doesn't predict such probabilities. We are one of the instances of $D$ but the theory doesn't  say which one. Indeed, it has no notion of `we'. 

To connect first person probabilities for what we observe with third person probabilities of what the universe exhibits a further assumption is needed. This assumption --- called a xerographic distribution 
\cite{HS09} --- specifies the probability that we are any one of the instances of $D$. The simplest and least informative assumption is that we are equally likely to be any one of the instances of $D$ that the universe exhibits. Put differently, it is the assumption that we are typical of those instances. This assumption is made throughout this paper\footnote{However sometimes assumptions of {\it a}typicality yield more predictive theories \cite{HS07,HS09}.}.

First person probabilities for what we observe are necessarily conditioned on the existence of at least one instance of our data $D$ in the universe --- us! Thus we write $p^\1p(z|\aoi)$ for the probability that we observe $z$. To calculate this, first calculate the joint probability $p^\1p(z,\aoi)$ as follows: Suppose the universe exhibits $n$ instances of $D$. Use an index $A$ running from $1$ to $n$ to distinguish these. Assuming typicality the xerographic distribution is $\xi_A = 1/n$.  Multiply this by the probability \eqref{prob_n}  for $n$ instances and sum over $A$. Finally sum over the number of instances from $n=1$ (at least one instance) to $n=N_h(z)$.  The factor of $n$ from the sum over $A$ cancels with the xerographic distribution to give
\begin{subequations}
\begin{align}
p^{\1p}(z, \aoi)= &\sum_{n=1}^{N_h(z)}\binom{N_h(z)}{n} \left(p_E(D)\right)^n\left(1-p_E(D)\right)^{N_h(z)-n} p(z) , \\
=& [1-(1-p_E(D))^{N_h(z)} ] p(z).
\label{probobs1}
\end{align}
\end{subequations}
The conditional probability $p^\1p(z|\aoi)$  is this joint probability divided by the probability just for $\aoi$:
\begin{equation}
\label{firstperson}
p^\1p(z|\aoi) = \frac{[1-(1-p_E(D))^{N_h(z)} ] p(z)}{\int dz [1-(1-p_E(D))^{N_h(z)} ] p(z)} .
\end{equation}
This is the probability that we observe $z$, and it is exactly the same as \eqref{thirdperson} derived with the aid of the symmetry.

\section{Quantum and Classical NBWF Fluctuations}
\label{classquantfluct}

The NBWF $\Psi$ is defined on the superspace of three-geometries and spatial matter field configurations. Here, we consider minisuperspace models defined by linearized perturbations away from closed, homogeneous and isotropic three-geometries and field configurations. Minisuperspace is spanned by the scale factor $b$ of the homogeneous three-geometries, the homogeneous value of the scalar field $\chi$  and the parameters defining the modes of  perturbation. We denote the latter collectively by $z=(z_1,z_2,...)$ and define these precisely in Section \ref{bottomup}. Thus, $\Psi=\Psi(b,\chi, z)$  

The NBWF is an integral of the exponential of minus the Euclidean action $I$ over complex  four-geometries and field configurations that are regular on a four-disk with a three-sphere boundary on which the four-dimensional histories take the real values $(b,\chi,z)$ \cite{HH83,Haw84}. Schematically we can write
\begin{equation}  
\Psi(b,\chi,z) =  \int_{\cal C} \delta a \delta \phi \delta\zeta  \exp(-I[a(\tau),\phi(\tau),\zeta(\tau)]/\hbar) .
\label{nbwf}
\end{equation}
Here, $a(\tau)$ and $\phi(\tau)$ are (complex) histories of scale factor and scalar field defining a homogeneous, isotropic background. The quantities $\zeta(\tau)=(\zeta_1(\tau),\zeta_2(\tau), \cdots)$ denote histories of  modes of fluctuation  away from homogeneity and isotropy in both metric and matter field. $I[a(\tau),\phi(\tau),\zeta(\tau)]$ is the Euclidean action. The integral is over geometries and matter fields  that  are regular  on a disk with only one boundary at which $a(\tau)$, $\phi(\tau)$ and $\zeta(\tau)$ take the values $b$, $\chi$, and $z$. The integration is carried out along a suitable complex contour ${\cal C}$ which ensures the convergence of \eqref{nbwf} and the reality  of the result \cite{HH90}. 

We restrict to linear fluctuations when only up to quadratic terms  in $\zeta$ are retained in the action in \eqref{nbwf}:
\begin{equation}
I = I^{(0)}[a(\tau),\phi(\tau)] + I^\epb [a(\tau),\phi(\tau),\zeta(\tau)]  .
\label{pertaction}
\end{equation}
(There is no linear term for the models considered in this paper.)
Then  $I^{(0)}$ describes the homogeneous isotropic background  and $I^\epb$ describes the linear and quadratic perturbations away from that background.

Suppose that in some region of superspace the integral in \eqref{nbwf} over $a(\tau)$ and $\phi(\tau)$ defining the homogeneous background can be approximated by the method of steepest descents. Then the wave function $\Psi$ will be a sum of terms of the form 
\begin{equation}
\Psi(b,\chi,z) \approx \exp\{[-I^{\epa}_R(b,\chi) + i S^{\epa}(b,\chi)]/\hbar\} \psi(b,\chi,z),
\label{semiclassback}
\end{equation}
one such term for each history $(a(\tau),\phi(\tau))$ that extremizes the action $I^\epa$, matches $(b,\chi)$ at the boundary of the disk, and is regular elsewhere. For each contribution $I^{(0)}_R (b,\chi)$ is the real part of the action $I^\epa[a(\tau),\phi(\tau)]$ evaluated at the extremizing history and $-S^\epa(b,\chi)$ is the imaginary part. The wave function $\psi$ is defined by the remaining integral over $\zeta$
\begin{equation}  
\psi(b,\chi,z) \equiv  \int_{\cal C}\delta\zeta  \exp(-I^\epb[a(\tau),\phi(\tau),\zeta(\tau)]/\hbar) .
\label{qftwf}
\end{equation}

As we showed\footnote{And as we intend to show in more detail in \cite{HHrules}.} in \cite{HHH08b}, classical {\it Lorentzian} histories are predicted in regions of superspace where $S^\epa(b,\chi)$ varies rapidly when compared with $I^\epa(b,\chi)$. Specifically, then $\Psi$ predicts an ensemble of suitably coarse-grained Lorentzian  histories $(b(t),\chi(t))$ that with high probability lie along  the integral curves of $S^\epa(b,\chi)$. Their relative probabilities are given by $\exp[-2I_R(b,\chi)]$, which is preserved along each history \cite{HHH08b}. 

When evaluated on one of these classical histories the wave function \eqref{qftwf} becomes a function of time,
\begin{equation}
\psi(z,t) \equiv \psi(b(t),\chi(t),z).
\label{wfot}
\end{equation} 
As shown in a variety of ways \cite{seqn}  the Wheeler-DeWitt equation implies a Schr\"odinger equation for $\psi(z,t)$
\begin{equation}
i\hbar d\psi(z,t)/dt = H(t) \psi(z,t)  . 
\label{seqn}
\end{equation}
The time dependent Hamiltonian describes the evolution of the state of the fluctuations in the background $(b(t),\chi(t))$. An inner product is induced from the generalized quantum mechanics on the full superspace \cite{Hartle95}. Equation and product define the quantum mechanics of the  fluctuation field $z$ in the homogeneous, isotropic background. 

In this way, the fluctuation fields can be thought of as quantum fields on the possible background classical spacetimes. The state of the fields is determined by the NBWF through \eqref{qftwf}. There is no independent assumption of a ``vacuum'' state. However, the Euclidean integral defining the NBWF is analogous to the Euclidean integral defining the ground state. It is therefore reasonable to expect the NBWF to imply that fluctuations are in something like a quantum field theory ground state early in the universe. This was shown explicitly in \cite{Hawking85} and we will show it explicitly for our model in the next  section\footnote{This also opens the possibility that the NBWF can predict corrections to popular assumptions about the vacuum of the fluctuation fields.}. Hence the NBWF provides a unified treatment of both classical homogeneous and isotropic backgrounds and the quantum fluctuations away from them. 

The integral defining the wave function in \eqref{qftwf} may itself be approximated by the method of steepest descents. Indeed, since the action is quadratic in $\zeta$ we expect that it can be evaluated exactly when the measure is suitable. Either way, the result for a particular extremum $a(\tau), \phi(\tau)$ of $I^\epa$ is
\begin{equation}
\psi(b,\chi,z) = A^\epb(b,\chi)  \exp\{[-I^{\epb}_R(b,\chi,z) + i S^{\epb}(b,\chi,z)]/\hbar\} .
\label{semiclassfluct}
\end{equation}
The extremizing history $\zeta(\tau)$ is regular on the manifold of integration and matches $z$ at its one boundary. 
$I^{\epb}_R(b,\chi,z)$ and $-S^{\epb}(b,\chi,z)$ are the real and imaginary parts of the action $I^\epb$ evaluated on this history and $A^\epb$ is a prefactor.

This fully quantum mechanical theory of fluctuations around a classical background universe will predict their classical behavior in  regions of superspace where $S^\epb(b(t),\chi(t),z)$ varies rapidly in $z$ compared to $I^\epb(b(t),\chi(t),z)$. The detailed conditions for this are called the `classicality conditions' \cite{HHH08b}. Specifically when they are satisfied the wave function \eqref{semiclassfluct} predicts an ensemble of suitably coarse grained, classical, Lorentzian histories $z(t)$ that with high probability lie along the integral curves of $S^\epb(b(t),\chi(t),z)$. The probabilities of the classical fluctuations in a given homo/iso background are then proportional to  $\exp[-2 I^{(2)}[b(t),\chi(t),z(t)]$.  In general we can expect the regions of superspace where perturbation modes behave classically to be different for different modes\footnote{In inflationary cosmology it is sometimes said that the modes `become classical' at a certain time as though there were a transition between quantum and classical physics. This is incorrect. Classical physics is not an alternative to quantum theory; it is an approximation to it. The modes are always quantum mechanical but a classical approximation only holds in certain regimes. It would be better to say that the modes enter a region where a classical approximation holds with a suitable coarse graining. But we will use the less accurate terminology with this understanding.} and we will see this in detail in what follows. 

With these techniques we will be able to treat both the quantum mechanics of fluctuations of the universe and their classical approximation.

\section{Bottom-Up Probabilities for Perturbations}
\label{bottom-up}\label{bottomup}

In this section we describe the calculation of the bottom-up probabilities for alternative four-dimensional classical histories of the universe that include linear fluctuations away from homogeneity and isotropy. These are the probabilities for classical behavior conditioned on the NBWF alone. They are the input to the calculation of top-down probabilities for observation described in the next section.

\subsection{Homogeneous Isotropic Histories}

We first review the bottom-up probabilities of the homogeneous isotropic histories predicted by the semiclassical NBWF \eqref{semiclassback}. These were calculated in \cite{HHH08a,HHH08b}, in a simple model consisting of a single scalar field moving in a quadratic potential. It was found that there is a one-parameter family of extremizing complex histories -- fuzzy instantons -- which obey the classicality conditions at the boundary where one evaluates the wave function and therefore predict a Lorentzian history. The different histories can be labeled by the magnitude of the complex scalar field $\phi_0\equiv |\phi(0)|$ at the `South Pole' (SP) of the corresponding fuzzy instanton. It was found \cite{HHH08a} that the classicality conditions require $\phi_0 \geq \phi_0^c \approx 1$. The relative probabilities of the different histories are given by $\exp[-2I_R(\phi_0)]$, where $I_R(\phi_0)$ is the real part of the Euclidean action of the fuzzy instanton.

A striking feature of the ensemble of classical histories in this model is the close connection it reveals between classicality and inflation \cite{HHH08b}. Specifically the histories have values of $H \equiv(db/dt)/{b}$ and $\chi$, which {\it all} lie within a very narrow band around $H = m \chi$ characteristic of Lorentzian slow roll inflationary solutions. It follows that a classical, homogeneous and isotropic universe {\it must have} an early inflationary state if the universe is in the no-boundary state. 

For sufficiently large $\phi_0$ there is an approximate analytic solution \cite{Lyo92} for the fuzzy instanton, 
\be
\label{ansol}
\phi (\tau)  \approx \phi(0) +i \frac{m \tau}{3},  \qquad a (\tau)  \approx \frac{i}{2m \phi (0)} e^{-im \phi(0)\tau+m^2 \tau^2/6}.
\ee
These solutions are the complex analogs of the standard `slow roll' inflationary solutions. They are valid in the region of the complex $\tau=x+it$ plane where $t$ is not so large that the slow roll assumption breaks down, and where $|a(\tau)| \gg 1$ so that the spatial curvature is exponentially negligible\footnote{The constant multiplicative normalization of the scale factor is determined by matching these solutions to the `no-roll' solutions 
$\phi(\tau) \approx \phi (0), \ a(\tau)  \approx \sin [m \phi (0) \tau]/m \phi (0)$ that are regular at the origin.}.
By tuning the phase of $\p0$ at the SP so that $ {\rm Im}[\phi(0)]= - \pi/6 {\rm Re}[\phi(0)]$ vertical lines given by $\tau = \pi/2 m  {\rm Re}[\phi(0)] +it $ are obtained along which both $a$ and $\phi$ are approximately real and describe Lorentzian inflating universes with the scalar field approximately equal to $\phi_0$ at the start of inflation.

The real part of the action of the fuzzy instantons in this approximation is
\begin{equation}
I_R (\p0)\approx -\frac{\pi}{2(m\p0)^2} \approx -\frac{2\pi}{(3m^2N(\p0))}
\label{action}
\end{equation}
where $N(\p0) \approx 3 \phi_0^2/2$ is the number of inflationary efolds in the classical history labeled $\p0$. Hence the bottom-up probabilities conditioned only on the NBWF are largest for classical histories with a small amount of inflation.

\subsection{Semiclassical Wave Function for Quantum Fluctuations}

Following the analysis of \cite{Hawking85,HLL93} we now calculate the wave function \eqref{qftwf} for linear scalar fluctuations  around the homogeneous isotropic histories predicted by the NBWF. We restrict attention to scalar perturbations, since these turn out to matter most for the top-down effects we are interested in here.  We write the perturbed metric as
\be
ds^2 = (1+2 \varphi) d\tau^2 + 2 a(\tau) B_{\vert i} dx^{i} d\tau + a^2(\tau) [ (1-2\psi) \gamma_{ij} +2E_{\vert ij}] dx^{i} dx^{j}
\label{pmet}
\ee
where $\gamma_{ij}$ is the metric of the unit radius three-sphere, $x^{i}$ are the coordinates on the three-sphere and a vertical bar denotes covariant differentiation with respect to $\gamma_{ij}$. 
Expanding the perturbations in the standard, normalized scalar harmonics $Q^{n}_{lm} (x^i)$ on $S^3$ gives the definitions
\be
\varphi = \frac{1}{\sqrt{6}} \sum_{nlm} g_{nlm} Q^{n}_{lm}, \qquad \psi =  \frac{-1}{\sqrt{6}} \sum_{nlm} (a_{nlm}+b_{nlm}) Q^{n}_{lm},
\ee
\be
B=  \frac{1}{\sqrt{6}} \sum_{nlm} \frac{k_{nlm} Q^{n}_{lm}}{(n^2-1)}, \qquad E= \frac{1}{\sqrt{6}} \sum_{nlm} \frac{3b_{nlm} Q^{n}_{lm}}{(n^2-1)}
\label{metrexp}
\ee
and the scalar field perturbation
\be
\delta \phi (\tau, x) = \frac{1}{\sqrt{6}}  \sum_{nlm} f_{nlm} Q^{n}_{lm}.
\label{scalarexp}
\ee
From here onwards we denote the labels $n,\ l,\ m$ collectively by $(n)$. The expansion coefficients $a_{(n)}, b_{(n)}, f_{(n)}, g_{(n)},k_{(n)}$ are functions of time only.

From the above expansions we see there are five scalar degrees of freedom. However, the functions $g_{(n)}$ and $k_{(n)}$ appear as Lagrange multipliers in the action. Variations of the action with respect to $g_{(n)}$ and $k_{(n)}$ result in the linear Hamiltonian and momentum constraints. In quantum cosmology the  NBWF satisfies the operator forms of these constraints \cite{HH91}. The wave function therefore depends only on the background variables $b$ and $\chi$ and on a single linear combination of the (boundary values of the) perturbation variables $a_{(n)}, b_{(n)}, f_{(n)}$ -- the three functions that describe the perturbed three geometry. One can take this linear combination to be the following (Appendix A),
\be
\zeta_{(n)} = a_{(n)}+b_{(n)} - \frac{H_E }{\dot \phi} f_{(n)}
\label{zet}
\ee
where $H_E \equiv \dot a/a$ and the subscript $E$ refers to quantities constructed with Euclidean time. Hence one has $\psi (b,\chi,z)$, where $z \equiv (z_{(1)},z_{(2)},...)$ are the real values of $\zeta=(\zeta_{(1)},\zeta_{(2)},...)$ at the boundary.
The variables $z$ are invariant under linear gauge transformations and approximately conserved outside the horizon \cite{Bardeen80,Weinberg08}. 

The wave function $\psi (b,\chi,z)$ can be found explicitly in the semiclassical approximation. To first order in perturbation theory \eqref{semiclassfluct} takes the form
\be
\psi (b,\chi,z) = \prod_{(n)} \psi_{(n)} (b,\chi,z_{(n)})   . 
\label{swave}
\ee
The action $I^{(2)}_{(n)}[b,\chi,z_{(n)}]$ of each mode is generally a {\it positive} quadratic function of $z_{(n)}$. Thus, in a regime where the perturbations are small and behave classically the bottom-up probabilities from \eqref{semiclassfluct} will favor vanishing perturbations and {\it homogeneous} classical histories.

An analytic approximation to the wave function \eqref{swave} was obtained in \cite{HLL93}, by solving the complex perturbation equations in the slow roll backgrounds \eqref{ansol}. In Appendix A we summarize this calculation and verify its accuracy by numerically calculating the perturbations around several representative members of the ensemble of exact complex extremizing geometries found in \cite{HHH08a,HHH08b}. We concentrate on perturbation modes that leave the Hubble radius during inflation. As we will see, these are the modes that are amplified by the time-dependent background, become classical and, ultimately, lead to the large-scale structures we observe today. 

The no-boundary condition of regularity at the SP requires $f_{(n)}$ and $a_{(n)}$ to vanish there. If $\tau \rightarrow 0$ labels the SP then the field equations imply that to leading order in $\tau$ one has
$\zeta_{(n)} = \zeta_{(n)} (0) \tau^{n}$, where $\zeta_{(n)}(0)\equiv \vert \zeta_{(n)}(0) \vert  e^{i\theta} \equiv \zeta_{(n)0} e^{i\theta}$ is a complex constant. Its phase $\theta$ should be fine-tuned such that $\zeta_{(n)}$ is real at the boundary, and its amplitude $\zeta_{(n)0}$ is determined by the value of the boundary perturbation $z_{(n)}$. 

At small $\tau$ the modulus of the complex `wavelength' $a/n$ of a perturbation mode will be shorter than the horizon size since $|aH_E| \rightarrow 1$ when $\tau \rightarrow 0$. In this regime we find the complex solution for $\zeta_{(n)}$ oscillates and is independent of the nature of the potential. On the other hand we show in Appendix A that at larger $\tau$, when $n\ll |aH_E|$, the general perturbation solution is a combination of a constant and a decaying mode. Hence one expects the wave function $\psi_{(n)} (b,\chi,z_{(n)})$ depends only on the behavior of the potential for values of $\phi$ near the value taken by $\phi (\tau)$ at the time the perturbation leaves the horizon. At horizon crossing the perturbation
$\zeta_{(n)}$ generally has an imaginary component. The requirement that $\zeta_{(n)}$ be real at the boundary therefore  means  that the phase $\theta$ of $\zeta_{(n)} (0)$ at the SP should be tuned such that the imaginary component of the subhorizon mode function matches onto the decaying mode when the perturbation leaves the horizon. It turns out that this implies that a perturbation mode will become classical when its physical wavelength becomes much larger than the Hubble radius, as is evident from Fig \ref{classicality} in Appendix A.
 
As a consequence of the decay of the imaginary component of the perturbation the real part of the Euclidean action $I^{(2)}_{(n)}(b,\chi,z_{(n)})$ tends to a constant when the mode leaves the horizon. This determines the bottom-up probabilities of the different classical perturbed histories predicted by the NBWF. Substituting the perturbation solutions \eqref{pertsol1} in the action \eqref{pertact} and normalizing one obtains, for all wavenumbers $n < \exp(3\phi_0^2/2) $, 
\be
\label{conserved}
p(z_{(n)}\vert \phi_0) \approx  \sqrt{\frac{\epsilon_{*}n^3 }{2 \pi H^2_{*}}}\exp \left[-\frac{\epsilon_{*}}{2H^2_{*}} n^3 z_{(n)}^2 \right]
\ee
where $\epsilon \equiv \dot \chi^2/H^2$ is the usual slow-roll parameter. The subscript $*$ on a quantity in \eqref{conserved} means it is evaluated at horizon crossing during inflation. Equation \eqref{conserved} specifies the bottom-up probabilities of linear, classical perturbations around the homogeneous isotropic histories predicted by the NBWF. One sees the probabilities of $z_{(n)}n^3$ are Gaussian, with variance $H^2_{*}/\epsilon_{*}$ characteristic of inflationary perturbations. 

Although \eqref{conserved} was derived using the slow roll approximation for the fuzzy instantons, we have numerically verified (Appendix A) that this result is accurate over most of the range of $\phi_0$ except near its lower bound $\phi_0^c$,
and for all modes that become classical except those that leave the horizon towards the very end of inflation\footnote{These corrections to \eqref{conserved} can in principle be calculated systematically, opening up the way to study the small deviations from the Bunch-Davis vacuum implied by the NBWF as discussed in Section \ref{classquantfluct}.}.

\subsection{Perturbed Classical Histories}

The evolution of perturbations in a classical background universe $(b(t),\chi(t))$ is in general given by a Schr\"odinger equation \eqref{seqn}. However in regions of superspace where $S^\epb(b(t),\chi(t),z)$ varies rapidly in $z$ compared to $I^\epb(b(t),\chi(t),z)$ the semiclassical wave function \eqref{swave} predicts an ensemble of suitably coarse grained, {\it classical}, Lorentzian histories $z(t)$ that with high probability lie along the integral curves of $S^\epb(b(t),\chi(t),z)$. Their relative probabilities are given by $\eqref{conserved}$, which is preserved along each history \cite{HHH08b}. 

We have seen that in inflationary histories, perturbation modes behave classically when their physical wavelength is larger than the Hubble radius. Since the modes that left the horizon during inflation are responsible for the large-scale structure we observe today, it is appropriate to evaluate the wave function of perturbations on a surface towards the end of inflation and to coarse-grain over modes that are inside the horizon at that time\footnote{The histories obtained by evolving forward Cauchy data taken at an earlier time, involving fewer modes, can be viewed as a coarse-graining of these.}.
The values of the perturbation modes at the boundary, together with their derivatives, provide Cauchy data for their
future classical evolution\footnote{The evolution of perturbations backwards in time, towards the initial singularity or the bounce \cite{HHH08b}, is generally not classical everywhere and can be obtained using \eqref{seqn}. This will be discussed elsewhere.}. 
The members of the classical ensemble of perturbation histories obtained in this way can be labeled by $(\phi_0, \zeta_{0})$.

Just after inflation ends the general solution for classical, long-wavelength ($n \ll b H$) perturbations (see e.g.\cite{Weinberg08}) implies the scalar metric perturbations remain essentially constant, with a small oscillatory component due to the oscillations of the background scalar field. The matter perturbation starts oscillating again when the Hubble radius becomes larger than the scalar field Compton wavelength $\sim 1/mb$. This behavior can also be seen in Fig \ref{pertmetric} (for $m^2=.05$), where inflation ends around $y \sim {\cal O}(60)$.
In realistic models the energy of the inflaton is then converted in ordinary matter and radiation, reheating the universe. Hence the solutions for the scalar matter and metric perturbations are not directly related to observations of the present universe. Fortunately, reheating occurs when the perturbation modes that are relevant for current observations are well outside the horizon, so that the variable $z_{(n)}$ is conserved. Hence the wave function $\psi (z,t)$ provides initial conditions for the classical evolution of perturbation modes after they re-enter the horizon at late times. 

To first approximation reheating takes place at a definite value of $\chi$. Hence one expects surfaces of constant scalar field during inflation to evolve to surfaces of constant temperature after reheating. The surface of last scattering will be such a surface. Variations in the observed temperature of the CMB arise e.g. from variations in the gravitational redshift of the surface of last scattering in different directions of observation, which are themselves determined by the perturbation $z(t)$. Quantities of particular interest in cosmology are averages over a particular pattern of perturbations at the surface of last scattering. The simplest examples are the multipole coefficients $C_l$ that characterize the average of a product of two temperature fluctuations in two different directions. 
Expressed in terms of $z_{(n)}$ the $C_l$'s involve a sum over the wavenumber $n$. However for $l \gg1$ the dominant contribution to this sum comes from perturbations with wavenumber $\bar n \approx l/r_L$, where $r_L$ is the radial distance from us to the surface of last scattering in the Robertson-Walker geometry (see e.g. \cite{Weinberg08}). 
This means there is a direct relation between the $C_l$'s and the variance of the probability distributions \eqref{conserved}. In particular CMB correlations on a certain angular scale at the present time provide information about the inflaton potential at the time of horizon exit of the relevant modes during inflation. For $10 \leq l \leq 50$ the CMB anisotropies are dominated by the Sachs-Wolfe effect. In this range the $C_l$'s are to a good approximation given by \cite{Weinberg08}
\be
C_l \approx \langle z_{(n)}^2 \rangle n^3  = \frac{8\pi^2 T_0^2 H^2_{*}}{9\epsilon_{*} l (l+1)}
\label{metricpt}
\ee
where $T_0^2$ is the present mean value of the temperature of the CMB. In the model we have considered $H^2_{*}/\epsilon_{*} \approx m^2 \chi_{*}^4$. Since galactic scales correspond to $\chi_{*}^2 \sim {\cal O}(50)$ and since observations require the gravitational potential to be $\sim 10^{-5}$ on these scales, the mass of the scalar field should be about $\sim 10^{-6}$ in Planck units. Larger scales leave the horizon earlier during inflation. During inflation one has $\chi_{*} \sim \ln (b_e/b_{*}) \sim \ln (\lambda_{ph}(n) H_{*})$ where $b_e$ is the scale factor at the end of inflation and $\lambda_{ph}=b/n$. Since $H$ is approximately constant during inflation this leads to a slightly red spectrum. Whereas this is a small effect on the range of currently observable scales, this has significant consequences on very large scales as we discuss below.

\section{Top-Down Probabilities for Perturbations}
\label{topdown}\label{top-down}

In this section we calculate the top-down probabilities $p(z_{(n)}|\aoi)$ for perturbation modes $z_{(n)}$ that are relevant for observation of the CMB. The Gaussian bottom-up probabilities \eqref{conserved} are an input to this calculation. Our particular aim is to determine in what models and under what conditions the top-down corrections can lead to observable
effects. We will find that this may be the case when the potential is such as to not allow a regime of eternal inflation.

We begin by reviewing the connection between bottom-up and top-down probabilities. This is the same as the connection derived in Section \ref{sec2} but written out here with the full machinery necessary to describe perturbations.

\subsection{Top-Down from Bottom-Up}

From the bottom-up probabilities \eqref{conserved} we seek to construct the (top-down) probabilities $p(z|D)$ for the present amplitudes of fluctuation observables  $z=(z_1,z_2,\cdots)$ conditioned on a subset $D$ of our total data. Suppose that $D$ can be divided into two parts: First, a part $D_s$ consisting of large scale observations that place the data $D$ on one or more surfaces of homogeneity $t_i(D_s,\phi_0)$ in each classical spacetime.  Observations of the present Hubble constant $H_0$  and local average energy density are an example. For simplicity we restrict attention to a single surface that we denote by $t$. The generalization to more is straightforward. 

The second part, $D_h$, consists of local observations that are largely independent of the large scale features of the spacetimes. Thus $D=(D_s,D_h)$. For each $\phi_0$ divide the surface labeled by $D_s$ into Hubble volumes and denote their total number by $N_h(D_s,\phi_0,\z0)$. Finally, denote by $p_E(D)$ the probability that the data $D$ occur in any one of the Hubble volumes on the surface $t$ and assume that  the probability of more than one occurrence in any one volume is negligible.  We can now follow the model in Section \ref{observations} to derive the \td probabilities for our observations of fluctuations. 

All we know from our local observations is that there is at least one occurrence of $D_h$ (abbreviated $\aoi_h$)  in one of the Hubble volumes (ours). The probability that there is at least one instance of $D_h$ in the classical spacetime labeled by $(\p0,\z0)$ is [cf. \eqref{atleast1}]
\begin{equation}
p(\aoh|D_s,\phi_0,\z0) =  1-[1-p_E(D)]^{N_h(t,\phi_0,\z0)} . 
\label{atleastone}
\end{equation}

Neither $\p0$ or $\z0$ is directly observable. But  in each classical spacetime we can determine the values of $z$ on the surfaces specified by $D_s$: $z=z(D_s,t,\p0,\z0)$.  Conversely, given $z$ and $\p0$ we can determine\footnote{To compress the notation we will not always write out the dependence of $\zs$ on $D_s$ and $\p0$.}  the amplitude of the fluctuations at the South Pole $\zeta_s(z)\equiv \z0(z,D_s,\p0)$ necessary to produce $z$ on the surface $t$. 
Thus we can write for the (top-down) probabilities $p(z|\aoi)$ 
\begin{equation}
p(z|\aoi)=\int d\p0 p(\p0,\zs)|\aoi) .
\label{2-1}
\end{equation}
This can be cast into a more useable form by using the joint probability [cf \eqref{reverse}]
\begin{subequations}
\label{2-2}
\begin{align}
p(\p0,\z0,\aoh|D_s) &=p(\aoh |D_s,\p0,\z0)p(\p0,\z0 |D_s) \\
                             &=p(\aoh|D_s,\p0,\z0)p(\z0|D_s,\p0)p(\p0|D_s) \label{2-2b}
\end{align}
\end{subequations}
Combining \eqref{2-1}, \eqref{2-2}, and \eqref{atleastone} we find the following formula for the top-down probabilities for fluctuations given at least one instance of the data $D$ [cf \eqref{thirdperson}] 
\begin{equation}
p(z|\aoi) \approx \frac{  \int d\p0 \ p(\zs|D_s,\p0)\{1-[1-p_E(D)]^{N_h(D_s,\p0,\zeta_s (z))}\}p(\p0|D_s)}
{  \int d\p0 d\z0 \ p(\z0|D_s,\p0)\{1-[1-p_E(D)]^{N_h(D_s,\p0,\z0)}\}p(\p0|D_s)}
\label{result}
\end{equation}
In \eqref{result} we expect the  dependence of the probabilities $p(\z0|D_s,\p0)$ and $p(\p0|D_s)$ on $D_s$ to be weak. They will be approximately proportional to $p(\z0|\p0)$ and $p(\p0)$ respectively except when the spacetime specified by $\p0$ does not contain a surface with data $D_s$. Then they are proportional to zero. 

The probabilities $p(z|\aoi)$ are for the values of the fluctuations the universe may exhibit given $\aoi$. (In the language of Section \ref{thirdtofirst} they are third person probabilities.) But we are interested in the (first person) probabilities for fluctuations in a particular history and inside our Hubble volume, where our specific instance of $D$ is located. For each $\phi_0$,  the underlying homogeneity is a symmetry that means that all predictions for observation will be the same in all Hubble volumes. The probability for any quantities derived from the $z$'s in our Hubble volume is the same as that derived from $p(z|\aoi)$ for any Hubble volume.

\subsection{Non-Gaussianity from Volume Weighting}
\label{nongaussvolwgt}

We now evaluate the TD probabilities $p(z_{(\bar n)} \vert \aoi)$ for different values of perturbation modes 
$z_{(\bar n)}$ in the classical ensemble of homo/iso histories with linearized perturbations. We are interested in particular in the modes that contribute to the CMB.  As reviewed at the end of the last section these are modes with approximately the same wavenumber that left the horizon the same number of efolds before the end of inflation in all members of the ensemble.
We denote the value of the relevant wavenumber in each history by $\bar n $. This depends on the duration of inflation and therefore on $\phi_0$. In terms of the angular scale this is given by $\bar n \approx l/r_L$, where $r_L$ is the radial distance to the surface of last scattering in the Robertson-Walker geometry (see e.g. \cite{Weinberg08}). To calculate the top-down probability $p(z_{(\bar n)}|\aoi)$ for the CMB relevant modes requires summing (coarse-graining) \eqref{result} over all other modes. 

As before we assume part of our data locate us on a surface of constant density in each member of the classical ensemble. 
The top-down probabilities \eqref{result} then involve the volume $N_h$ of this surface. This is most easily calculated in the $f_{(n)}=b_{(n)}=0$ gauge, where surfaces of constant density are constant time surfaces with volume [cf.\eqref{pmet}]
\be
V = V_0 + \delta V = b^3 \int d^3 x \sqrt{\gamma} (1-2\psi)^{3/2}.
\label{vol}
\ee
The leading correction to $V_0$ averages to zero over the surface, but the second order term leads to a change in volume.
In terms of the gauge invariant variable $z_{(\bar n)}$ one has $\delta V/V_0 = \sum_{(n)} z_{(n)}^2/8\pi^2$. 
Hence the number of present Hubble volumes in the different histories of the ensemble is given by 
\be \label{volume}
N_h  (D_s, \phi_0,z) = N_h^0 (D_s,\phi_0) \bigg(1+\sum_{(n)} \frac{z^2_{(n)}}{8\pi^2}\bigg)\exp \left(\frac{9}{2} \phi_0^2\right),
\ee
where $N_h^0 (D_s,\phi_0)$ varies slowly with $\phi_0$ and depends on the present Hubble constant, the details of reheating etc. The range of $n$ in the sum encompasses all modes that left the horizon during inflation  and are therefore classical.  Its upper limit $n_m$ therefore depends on $\phi_0$ and is approximately given by $n_m \approx \exp\left(3\phi_0^2/2\right)$. Using the BU distribution \eqref{conserved} of $z_n$ the expected value of the sum in \eqref{volume} can be bounded by the variance of the longest wavelength perturbations in each history --- with $n=n_m$ --- yielding $\langle \sum_{(n)} z^2_{(n)} \rangle  \leq m^2 \phi_0^4$.

\begin{widetext}
The TD distribution is of the form \eqref{result}. Using the analytic approximations \eqref{action} and \eqref{conserved} of the BU probabilities of homo/iso histories with linearized perturbations one finds {for the top-down probability of the CMB relevant modes}\footnote{In \eqref{td} we have not taken in account the Jacobian that arises when one changes the integration measure from $d\zeta_{(n)0}$ to $dz_n$, because this is polynomial in $\phi_0$ (see Appendix and also \cite{HLL93}) and therefore hardly affects the TD probabilities.}
\be 
p(z_{(\bar n)} |\aoi)  \propto 
\int d\phi_0 \left[\prod_{(n)\neq (\bar n)} d\zeta_{(n)0} 
\exp \left(-\frac{z_{(n)}^2}{2\sigma_n^2} \right) \right]
\left[1-(1-p_E)^{N_h}\right)] 
\exp \left(-\frac{z_{(\bar n)}^2}{ 2\sigma_{\bar n}^2} \right) 
\exp \left( \frac{4\pi}{3m^2 N} \right).
\label{td}
\ee
Here $\sigma_n^2 (\phi_0) \equiv H^2_{*}/2 \epsilon_{*} n^3$ and the product is taken over all wavenumbers  $n$ up to $n_m$.
The integrals over $\zeta_{(n)0}$ can be evaluated analytically without further approximations. This yields
\beq 
p(z_{(\bar n)} |\aoi) &  \propto & 
\int d\phi_0  \left[ \prod_{(n)\neq (\bar n)}   \sqrt{2\pi \sigma_n^2} - 
(1-p_E)^{\bar N_h}  \prod_{(n)\neq (\bar n)} 
\frac{\sqrt{2\pi \sigma_n^2}}{\sqrt{1-(\sigma_n^2/4\pi^2) N_h^0 e^{3N} \log (1-p_E)}} \right] \nonumber\\
& & \qquad \times
\exp \left(-\frac{z_{(\bar n)}^2}{ 2\sigma_{\bar n}^2}  \right) 
\exp \left(\frac{4\pi}{3m^2 N} \right) 
\label{tdpert}
\eeq
where $\bar N_h \equiv N_h^0 (N) (1+z^2_{(\bar n)}/8\pi^2)\exp (3N)$.
\end{widetext}
In \cite{HHH08a,HH09} we have argued that for realistic values\footnote{That is, assuming we are a typical instance of $D$  and conditioning on our actual observational situation. Top-down probabilities conditioned on fewer or altogether different data can be calculated as well and may be of interest. When $p_E > 1/ N_h$, the first term in \eqref{tdpert} provides the dominant contribution to such TD probabilities which are therefore Gaussian} of $p_E$, volume weighting applies in the ensemble of  homogeneous isotropic histories even in models where the potential admits inflationary solutions all the way up to the Planck scale, corresponding to values $\phi_0^{pl} \sim 1/m$. This is because one can easily find data $D$ for which $p_E \ll 1/ N_h^{pl}$, where $\log (N_h^{pl}) \approx 3N (\phi_0^{pl}) = 9/2m^2 \approx  10^{12}$. In this regime the top-down factor reduces to $p_E N_h$ and the probability $p_E$ cancels out, as discussed in Section II. A single perturbation mode on currently observable scales hardly changes the volume $N_h$. Hence the factor $(1-p_E)^{\bar N_h}$ in \eqref{tdpert} is approximately given by $1-\bar N_h p_E $ for realistic values of $p_E$. 

The product in the second, non-Gaussian term in \eqref{tdpert} further simplifies in histories where
\be \label{rare2}
p_E < \bigg[\Big(\sum_{(n)} \sigma_n^2/4\pi^2\Big) N_h^0 e^{9\phi_0^2/2} \bigg]^{-1}.
\ee
When  $\phi_0 < 1/\sqrt{m}$ this condition automatically holds 
when the data are rare in the background history  because  the sum over $\sigma_n$ is smaller than one. In contrast, in eternally inflating histories this is a stronger condition than the requirement used above that the data be rare in the homo/iso background. Indeed, in histories with a regime of eternal inflation and hence $\phi_0 > 1/\sqrt{m}$ one finds $(\sum_{(n)} \sigma_n^2/8\pi^2) \approx m^2 \phi_0^4 \gg 1$, due to long wavelength perturbations that leave the horizon when $\chi(t) > 1/\sqrt{m}$. This reflects the fact that in eternal inflation, perturbations can significantly change the volume of surfaces of constant scalar field and therefore the possible locations where our data can be. However, based on the arguments in \cite{HH09} it appears plausible that the condition \eqref{rare2} holds {\vf with realistic values of $p_E(D)$ } even in eternally inflating histories. Hence the TD probabilities $p(z_{(\bar n)} |\aoi)$ are approximately given by
\begin{widetext}
\beq \label{tdpert2}
p(z_{(\bar n)} |\aoi) & \propto & \int d\phi_0 \left(\prod_{(n)\neq (\bar n)}  \sqrt{2\pi \sigma_n^2} \right) \left[  1 - 
\frac{1-p_E \bar N_h }{\sqrt{1+p_E (\sum_{(n)\neq (\bar n)} \sigma_n^2/4\pi^2) N_h^0 e^{3N}}} \right]\nonumber \\
& & 
\times \exp \left(-\frac{z_{(\bar n)}^2}{ 2\sigma_{\bar n}^2}  \right) 
\exp \left( \frac{4\pi}{3m^2 N} \right) 
\eeq
\end{widetext}
Expanding the square root and including the normalization factor in \eqref{result} yields
\be \label{tdpert3}
p(z_{(\bar n)} |\aoi)= \frac{\int d\phi_0 \left(\prod \sqrt{2\pi \sigma_n^2} \right)
N_h^0  \left(1+ \sum \frac{\sigma_n^2}{8\pi^2} +\frac{z_{(\bar n)}^2}{8\pi^2} \right)
\exp \left[-\frac{ z_{(\bar n)}^2}{ 2\sigma_{\bar n}^2}  \right] 
\exp \left[ 3N+\frac{4\pi}{3m^2 N} \right] }
{\int d\phi_0 \prod_{(n)}  \sqrt{2\pi \sigma_n^2}  
N_h^0  \left(1+\sum_{(n)} \sigma_n^2/8\pi^2 \right)
\exp \left[ 3N+\frac{4\pi}{3m^2 N } \right] }
\ee
where the product and sum in the numerator are taken over all classical modes except the mode labeled by $(\bar n)$.
The probability $p_E$ has cancelled out. In models {\it with a regime} of eternal inflation, the volume weighting $\exp(3N)$ implies  that the dominant contribution to the integrals in \eqref{tdpert3} comes from histories with the largest values\footnote{What these are depends on the model, i.e. where $V$ becomes too steep for inflation to occur. Below we assume for simplicity this only happens at the Planck scale, corresponding to $\phi_0^{pl} \approx 1/m$.} of $\phi_0$ and hence a long period of inflation. 
In histories of this kind $\sum_{(n)} \sigma_n^2/8\pi^2 \gg 1$. Hence the normalizing factor in the denominator makes the non-Gaussian TD corrections in $z_{(\bar n)}$ extremely small, yielding
\be \label{tdpertEI}
p(z_{(\bar n)} |\aoi) \approx p(z_{(\bar n)} |\aoi,\phi_o^{pl}) \approx \frac{1}{\sqrt{2\pi \sigma_n^2} } \exp \left(-\frac{\epsilon_{*}}{H^2_{*}} \bar n^3 z_{(\bar n)}^2 \right) 
\ee

{\vf Even in the context of quadratic potentials it is possible to construct models without a regime of eternal inflation for instance by restricting the physically allowed range of $\phi$. In such models, where all histories have $\phi_0 < 1/\sqrt{m}$, the integral over the other perturbation modes has little effect and the non-Gaussian TD corrections in $z_{(\bar n)}$ remain relevant in contrast to the result above.}
On the other hand, in this case the volume weighting does not significantly change the BU distribution of histories with different $\phi_0$ \cite{HH09}, so that the integral over $\phi_0$ is dominated by histories with the smallest amount of inflation compatible with the data $D$. The TD distribution in a background of this kind is approximately given by
\be \label{tdpertnoEI}
p(z_{(\bar n)} |\aoi) \approx \frac{1}{ \sqrt{2\pi \sigma_{\bar n}^2}} \frac{1+z_{(\bar n)}^2/8\pi^2 }{1+ \sigma_{\bar n}^2/8\pi^2} 
\exp \left(-\frac{\epsilon_{*}}{H^2_{*}} \bar n^3 z_{(\bar n)}^2 \right). 
\ee
Hence in models without a regime of eternal inflation the NBWF predicts we should observe a slightly non-Gaussian spectrum of perturbations even though their BU distribution is Gaussian.

It is possible to calculate the \bu probabilities for the fluctuations pertaining to the  CMB  by focussing only on the relevant modes and ignoring all others in a restricted  minisuperspace model. At the \bu level all modes are independent in the linear approximation.  However, we have seen here that this is not possible for the \td probabilities for CMB observations. The observations  may only probe the wavelengths characteristic of only a few modes, but the top-down weighting depends on all of them.  In a minisuperspace approximation consisting of homo/iso histories with {\it a single} perturbation mode $z_{(\bar n)}$, we would have predicted non-Gaussianity even in models of eternal inflation. When we include all modes in our analysis the answer is qualitatively different.  Quantum mechanics then instructs us to coarse-grain over perturbations we do not observe. In eternally inflating histories this reduces the non-Gaussianity as in eq \eqref{tdpert3}.

As discussed earlier, CMB temperature correlations on a given angular scale provide an excellent probe of the TD distribution for $z_{\bar n}$ especially at large $l$, where cosmic variance is limited and where the dominant contribution comes from modes with a particular wavenumber $n$. Hence the prediction of non-Gaussianity with a specific shape in models without eternal inflation leads to the possibility of determining whether or not eternal inflation took place. The fact that we can learn something about the global structure of the universe from local observations conditioned on local data $D$ can be traced to the quantum state which predicts non-local correlations. The predicted level of non-Gaussianity in models without a regime of eternal inflation is  small on currently observable scales. However even a small departure from a Gaussian spectrum may be detectable with future observations. We will therefore return in future work to a more detailed analysis of top-down corrections in the CMB anisotropies.

\section{Backreaction in the Regime of Eternal Inflation}
\label{topdown_inei}

The expected amplitude of long wavelength perturbations that leave the horizon in the regime of eternal inflation is large. Indeed, it follows from \eqref{conserved}  that $H^2_{*} \geq \epsilon_{*}$ when $\chi_{*} \geq \sqrt{m}$ and hence 
$\langle z_{(n)}^2 \rangle n^3 >1$. Since in models of eternal inflation histories with $\phi_0 > 1/\sqrt{m}$ dominate the TD probabilities \cite{HHH08a,HH09}, this means there is a significant probability for our universe to be strongly {\it inhomogeneous} on the largest scales in models of this kind.

This {\vf inhomogeneity}  has important implications for the possible locations of our data, because these {\vf typically}  confine us to one or several surfaces of constant density. A calculation in perturbation theory of the expected fractional change in the volume $V(t)$ of a surface of constant scalar field, due to combined effect of all fluctuations outside the horizon yields, from \eqref{vol} and using \eqref{metricpt},
\be
\langle \frac {\delta V}{V_0} (t)  \rangle = \frac{1}{8\pi^2}  \int^{n_m(t)} d^3 n \langle z_{n}^2 \rangle \approx \frac{1}{8\pi^2} \frac{H^2(t)}{\epsilon(t)} 
\label{bent}
\ee
where $n_m(t)=Hb(t)$ and $V_0(t)=2\pi^2 b^3(t)$ is the volume of a surface which is at time $t$ in the unperturbed geometry. Hence, {\vf for instance,  the expected volume of the reheating surface} in perturbed histories with $\phi_0 >1/\sqrt{m}$ can differ significantly from the reheating volume in the homogeneous isotropic background. This indicates perturbation theory may be inadequate to calculate the precise shape of the reheating surface in eternally inflating histories.
In fact, it has been argued (see e.g. \cite{EIrefs,Winitzki08,CREM}) -- albeit in part based on perturbation theory  -- that starting with a finite inflationary volume in the regime of eternal inflation, backreaction effects give rise to a significant probability for developing constant scalar field surfaces of arbitrarily large or even infinite volume\footnote{Numerical simulations of perturbed classical universes in this regime using stochastic techniques \cite{Star86,Linde96,Gratton05,Vanchurin00} provide some support for this.}.

This implies that in models of eternal inflation it may not be correct to assume that our data is rare in every history of the ensemble\footnote{We note however that the connection between the volume of the reheating surface and that of the surface of constant present matter density is rather complicated, since large-scale perturbations are large. This is a caveat in the analysis of top-down probabilities in this model.}. Instead in a subset of histories the more general weighting \eqref{result}, or even its common limit, may apply in the calculation of TD probabilities rather than volume weighting.

However this {\vf more general weighting} is unlikely to change our results for the TD probabilities $p(z_{(\bar n)} |\aoi)$ obtained in Section \ref{nongaussvolwgt}, as we now explain.
Let us assume, as before, that the data are rare in all background histories, i.e. $p_E \ll 1/N_h^{pl}$. The top-down weighting then implies that eternally inflating histories with large $\phi_0$ provide the dominant contribution to the TD distribution \cite{HHH08a}. Volume weighting will apply in approximately homogeneous and isotropic histories of this kind, yielding the Gaussian contribution to the TD distribution given in \eqref{tdpertEI}. However, if backreaction leads to a significant probability for the reheating surface to be infinite then the main contribution to the TD distribution will come from significantly perturbed histories where our data are common because $N_h$ is {\vf large or} infinite. But in such histories the TD weighting in \eqref{td} equals one. Hence predictions for observations are given by the bottom-up probabilities. One expects BU probabilities of observable fluctuations not to be affected by backreaction effects, since perturbation modes on currently observable scales leave the horizon well outside the regime of eternal inflation where these effects are negligible. Hence we expect the result \eqref{tdpertEI} remains unchanged when backreaction is taken in account.
 
Roughly speaking, one could say that in these models, by selecting histories with a large number of efolds, the top down weighting also makes it likely for there to be a Hubble volume with any given local perturbation on surfaces of constant density. Indeed in a sufficiently large universe anything will happen {\vf somewhere}. Hence the probability that a typical observer sees a particular fluctuation is determined by the relative frequency with which different fluctuations occur. But this is precisely what is given by the BU probabilities. This is an example where the quantum state specifies a measure for local prediction in cosmology that is well behaved for spatially large or infinite universes.

\section{Conclusion}
\label{conclu}

The approach of this paper to cosmology in the regime of eternal inflation is significantly different from many others \cite{measurerefs}. We have started from the fundamental assumption that the universe, including all its contents, is a closed quantum mechanical system. We have explored the consequences of this for prediction in the regime of eternal inflation in simplified models in the context of the low-energy approximate quantum theory of gravity. 
  
Like any other closed quantum system the universe has a quantum state. The NBWF is the model for this state used here. Bottom-up probabilities for the different, coarse-grained histories of the universe and its contents follow from this state and not from a further posited measure. Classical behavior of spacetime geometry is not assumed. Rather the ensemble of possible classical histories of the universe is derived from its quantum state. 
 
Observers are not assumed to necessarily exist, nor to be unique, nor to be essentially classical systems outside the reach of quantum mechanics. Rather they are quantum subsystems of the universe described by certain data with a probability to exist in any Hubble volume and a probability to be exactly replicated elsewhere in the universe.  

Probabilities relevant for observations are top-down probabilities that take in account the observing system as a quantum subsystem of the universe. The starting point for the calculation of top-down probabilities are the bottom up probabilities {\vf for four-dimensional histories} conditioned on just the NBWF  --- the universe sub specie aeternitatis.

The NBWF predicts a particular ensemble of classical, inflationary histories with a characteristic set of perturbations that emerge from quantum fluctuations. The bottom-up probabilities favor histories with a small number of efolds \cite{HHH08a}. 
The perturbations are Gaussian with variance $V(\chi)/ \epsilon$ evaluated at horizon crossing, where $\epsilon$ is the slow-roll parameter (eq. \eqref{conserved}). Therefore in histories with a regime where $V(\chi) > \epsilon$, significant probabilities are predicted for large fluctuations that left the horizon while this condition holds. This is {\vf called} the regime eternal inflation. The NBWF thus predicts that histories of this kind are {\it inhomogeneous} on the large scales that left the horizon during such a regime. In particular it predicts that any constant $\chi$ surface, such as the reheating surface, can differ significantly from the same surface in the homo/iso background. This result resonates well with other discussions of eternal inflation as well as numerical simulations using stochastic techniques \cite{Star86,Linde96,Gratton05,Vanchurin00}.

Top-down probabilities are constructed from bottom-up probabilities by further conditioning on some part of our data that includes a description of the observational situation within the universe. If one conditions on data $D$ that localize the observer on one or several surfaces in each history then the general weighting \eqref{result} connects top-down probabilities to bottom-up ones. This weighting is not a choice, or a postulate, or a proposal. Instead it arises necessarily from four considerations: 1) Our data $D$ occur within a given Hubble volume only with some quantum probability $p_E$.  2) In a large universe our data may occur elsewhere with significant probability. 3) All we know about the universe is that our history exhibits at least one instance of it. 4) {\vf An assumption}  that we are equally likely to be any of the instances of $D$ that our universe exhibits.

Volume weighting arises as an approximation to \eqref{result} {\it only} when our data are rare in all histories in the ensemble that are predicted with any significant probability. For realistic values of $p_E$ \cite{HH09} we find this implies that top-down probabilities favor histories with a large number of efolds in models that have a parameter regime where $V > \epsilon$, with $\epsilon$ the slow-roll parameter \cite{HHH08a,HHH08b}. Unlike this approximation, the general weighting \eqref{result} is well behaved even when spatial volumes become infinite. In fact for very large volumes, the quantum nature of the observational situation implies that the top-down probabilities for observations converge to the bottom-up probabilities\footnote{This resonates with \cite{Linde09} where it is argued that the total number of locally distinguishable FRW universes generated by eternal inflation is finite. Here we have seen that the top-down probabilities for different values of local perturbations become indistinguishable in very large universes.}. This is an important difference with other discussions of eternal inflation, usually not based on quantum cosmology. There the infinite volume limit instead leads to ambiguities. In those cases, {\vf  to predict the outcome of our observations unambiguously,  a measure must be introduced} that regularizes the infinitely large spatial volumes that arise in the regime of eternal inflation. By contrast, in quantum cosmology the wave function provides the only measure needed for unambiguous prediction\footnote{It would {\vf be of interest} to compare top-down probabilities calculated from the NBWF with the predictions of other measures employed in the study of eternal inflation.}. Furthermore, it does this as part of a unified framework that also explains the origin of inflation and of classical spacetime itself. 

In this paper we have calculated the top-down probabilities for different fluctuations in models with a single scalar field $\chi$ with a quadratic potential $m^2 \chi^2$. We find that the NBWF predicts a significantly inhomogeneous universe on very large scales and a Gaussian spectrum of small perturbations on currently observable scales when there is a regime of eternal inflation, i.e. $\chi > 1/\sqrt{m}$ in the early universe. The inclusion of backreaction effects of perturbations may give rise to histories with a truly infinite reheating surface, but we have no indications this leads to a breakdown of the calculational framework nor do we expect this to change this specific result.

By contrast, in models where the scalar field takes values only in a restricted range\footnote{Or models where quantum corrections render the potential too steep so that the regime of eternal inflation sets in only at the Planck scale.} that does not include a regime where $V > \epsilon$, we find the top-down probabilities predict large-scale homogeneity and a slightly non-Gaussian spectrum of observable fluctuations, for realistic values of $p_E$. The predicted level of non-Gaussianity is small on observable scales but potentially detectable with future experiments. More generally we expect it to be true that the top-down weighting leads to some non-Gaussianity only in models without eternal inflation, and therefore to the possibility to test whether our universe exhibits a regime of eternal inflation.

The differences between the TD and BU probabilities are striking.  Bottom up probabilities favor past inflation but only in small amounts. In contrast, top-down probabilities favor a large number of efolds of past inflation. Bottom up probabilities favor a homogeneous universe. Top-down probabilities predict a universe that is significantly inhomogeneous on scales much larger than the present horizon in models with eternal inflation. 
 
The top-down probabilities for prediction exemplified by \eqref{result} depend only on data $D$ within our past light cone. (In the present models this data is approximated by data on a spacelike surface in our Hubble volume.)  But they also depend on the implications of the theory for the structure of the universe on scales much larger than the present horizon. That is because top-down probabilities depend not only on what the data are on our past light cone, but also on where light cones with that data may be located in spacetime. This is determined in part by the quantum state, which predicts non-local correlations and in particular specifies what the allowed classical spacetimes are.

Turning this connection around we see that from local observations we may draw inferences about the structure of our universe outside the present horizon, assuming of course that the theoretical framework behind these predictions is secure.
The TD predictions for the spectrum of primordial perturbations provide a striking example of this. These predict a specific form of non-Gaussianity, but only in histories where we are rare. Any observation of this non-Gaussianity would therefore provide valuable information about the possible locations of our data and place an upper bound on the size of our universe. If by contrast this non-Gaussianity turns out to be absent in the perturbation spectrum this would be evidence 
for a much larger, eternally inflating and therefore possibly infinite universe. 

Thus if the values of the top-down probabilities depend on the large scale structure of the universe then the results of the observations they predict offer the opportunity to probe this structure. This striking connection between global structure and local observation is ultimately traceable to the NBWF which, like any quantum state, is defined globally not locally.

The underlying homo/iso symmetry has of course greatly simplified the calculation of TD probabilities in this paper.
The symmetry means all Hubble volumes on the surface where the data $D$ occur are equivalent, which essentially allows one to ignore all other instances of our data and focus on our own. In particular, the probabilities for different values of a perturbation mode $z_{(n)}$ in our own Hubble volume given $\aoi$ are the same as the probabilities that the universe exhibits different values of perturbations $z_{(n)}$ in any Hubble volume given $\aoi$. The symmetry therefore automatically `organizes' the different Hubble volumes, even when $D$ occurs on infinitely large surfaces. 

To apply the top-down approach to string theory which, it has been argued, at low energies predicts a potential landscape with finitely many vacua with different physics, one  must generalize the calculations in this paper to models where the possible locations of $D$ are not all connected by symmetry\footnote{The range of possible locations depends on the set of histories involved and therefore on the coarse-graining. The latter is in turn determined by the question one asks.}. In models of this kind, one expects the NBWF to select inflating histories that roll down from flat patches in the landscape where the slow roll conditions hold. However it appears plausible that besides histories where the background is homogeneous and isotropic, the ensemble also includes histories where our data occur on homogeneous surfaces in open FRW universes that are bubbles inside de Sitter space. This is because one can get histories of this kind from complex Coleman-De Luccia instantons that obey the no-boundary condition of regularity \cite{CdL80}. If one neglects collisions between bubbles then all locations inside bubbles of the same type are equivalent, and only one `representative' location enters in the calculation of TD probabilities. By contrast, the relative probability of finding our data in bubbles of different types (or in histories without bubbles) is important. In the NBWF this is given by the ratio of the real part of the actions of the corresponding instantons, yielding a well-defined prediction. Thus, at the present moment, we see no obstacle of theory or practice to extending the results of this paper to more general and realistic models of the implications of a quantum universe.

\vspace{.3cm}

\noindent{\bf Acknowledgments}  We thank Ben Freivogel, Jaume Garriga, Allan Guth, Ricardo Monteiro and Alex Vilenkin for helpful discussions. We also thank the Theory Division of CERN for its hospitality during part of this work. The work of JH was supported in part by the National Science Foundation under grant PHY05-55669.

\appendix
\section{Semiclassical Wave Function of Linear Perturbations}

In this Appendix we calculate the wave function of linearized perturbations around the homogeneous isotropic saddle point histories discussed in Section \ref{bottomup}A.

As discussed in Section \ref{bottomup}B, the wave function of linear perturbations depends on the background variables $b$ and $\chi$, and on a single linear combination of the perturbation variables $a_{(n)}, b_{(n)}$ and $f_{(n)}$ that describe the perturbed {\vf complex extremizing four} geometry. We work with the following gauge-invariant linear combination (see also \cite{Shirai88,HLL93}), 
\be
\tilde \zeta_{(n)} = a^3 [ \dot \phi (a_{(n)}+b_{(n)}) - H_E^{\ } f_{(n)}],
\label{zet1}
\ee
where $\tilde \zeta$ tends to a real value $\tilde z$ at the boundary. To calculate  $\Psi (b,\chi,\tilde z)$ it is convenient to return to the original perturbation variables \eqref{pmet} and to choose a particular gauge to find the solutions that extremize the action. One can then rewrite the result in terms of $\tilde z_n$ and therefore express the wave function in a gauge invariant way. (In Section \ref{bottomup} we have written the wave function in terms of $z=\tilde z/a^3 \dot \phi$, which is conserved outside the horizon and therefore closely related to physical (observable) quantities.)

A general linear scalar gauge transformation allows one to set $E=B=0$ in \eqref{pmet}, or  $b_{(n)}=k_{(n)}=0$ in terms of perturbation modes. This is the Newtonian gauge in which $g_{(n)}=-a_{(n)}$, and the equations that govern the fluctuations read \cite{HLL93}
\begin{subequations}
\label{eucperteqns}
\be
\ddot a_{(n)} +4H_E^{\ } \dot a_{(n)} - (3m^2 \phi^2 -2/a^2) a_{(n)} = -3 \dot \phi \dot f_{(n)} -3 m^2 \phi f_{(n)}
\ee
\be
\ddot f_{(n)} +3H_E^{\ } \dot f_{(n)} - (m^2 +(n^2-1)/a^2) f_{(n)} = -4 \dot \phi \dot a_{(n)} -2m^2 \phi a_{(n)}
\label{pert2eq}
\ee
\be
\dot a_{(n)} +H_E^{\ } a_{(n)} = -3 \dot \phi f_{(n)}
\label{pert3eq}
\ee
\end{subequations}
We consider a coarse-graining in which we concentrate on perturbation modes that leave the Hubble radius during inflation. These are the modes that get amplified by the time-dependent background and, ultimately, lead to the large-scale structures we observe today. 

The no-boundary condition selects solutions of \eqref{eucperteqns} that are regular at the SP.
This means $f_{(n)}$ and $a_{(n)}$ must vanish as $\tau \rightarrow 0$. From eqs \eqref{eucperteqns} and regularity of the background it follows that near the SP, the leading order term in $\tau$ is given by
\be \label{SPsol}
f_{(n)} = \zeta_{(n)} (0) \tau^{n-1}, \qquad a_{(n)} = - \frac{3m^2 \phi (0)}{4(n+2)} \zeta_{(n)}(0) \tau^{n+1}
\ee
where $\zeta_{(n)}(0)\equiv \vert  \zeta_{(n)}(0) \vert  e^{i\theta} \equiv  \zeta_{(n)0} e^{i\theta}$ is a complex constant. The phase $\theta$ should be fine-tuned such that $\zeta_{(n)}$ is real at the endpoint $\upsilon$. The amplitude $\zeta_{(n)0}$ in turn is determined by $z_{(n)}$. At the SP $\zeta_{(n)0}$ is thus a free parameter which can be used to label the different histories. The ensemble of perturbed histories can therefore be labeled by $(\phi_0, \zeta_{0})$. 

\begin{figure}[t]
\includegraphics[width=3.2in]{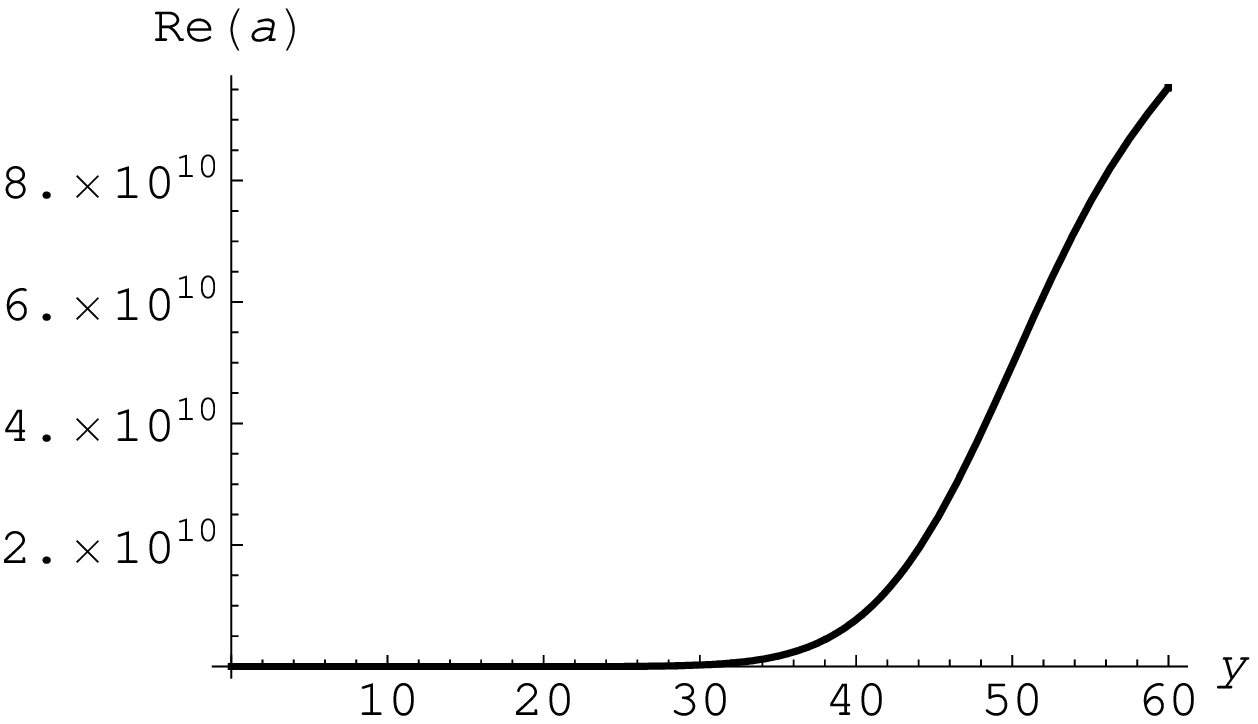} \hfill
\includegraphics[width=3.2in]{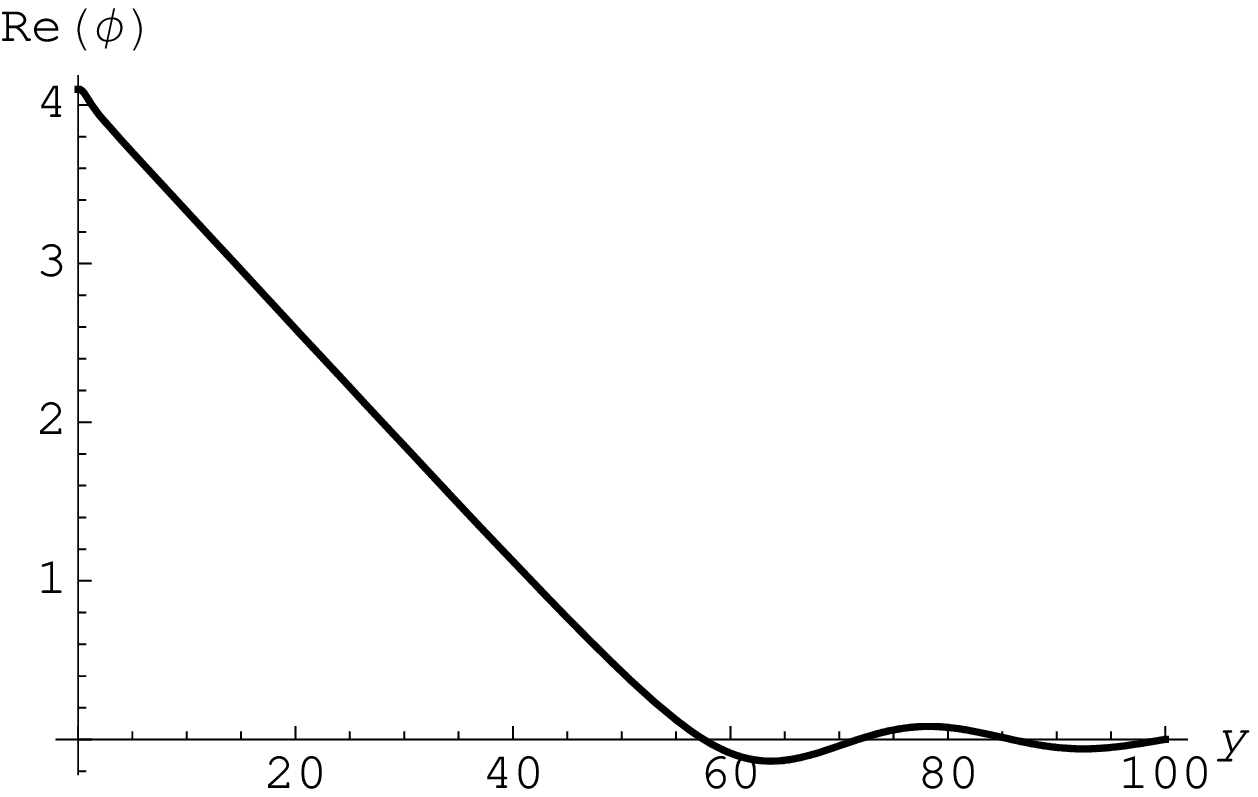} 
\caption{The real part of the scale factor (left) and the scalar field (right) of the complex homogeneous isotropic slow-roll solution labeled by $\phi_0=4$, with $m^2=.05$. This is shown here along the vertical part of a contour in the complex $\tau$-plane that first runs from the origin to $ X \approx \pi/2m\phi_R(0)$ and then upward along the $y$-axis. The turning point $X$ and the phase of $\phi(0)$ have been fine-tuned so that $a$ and $\phi$ tend to real functions along the vertical part of the contour. This happens very rapidly, so that the solution behaves classically already at $y \geq {\cal O} (1)$.}
\label{background}
\end{figure}

At early times, when the physical wavelength $a/n$ of the perturbation mode is smaller than the Hubble radius $H_E^{-1}$, the metric perturbation $a_{(n)}$ does not significantly affect the evolution of the matter perturbation $f_{(n)}$. Specifically the terms on the right-hand side in \eqref{pert2eq} are negligible in slow-roll backgrounds \eqref{ansol} when $n \gg |H_E^{\ }a |$, so that for $n \gg1$ the matter perturbation equation reduces to
\be
f_{(n)}''+2 {\cal H}_E^{\ } f_{(n)}' -(n^2-1) f_{(n)} =0.
\ee
Here prime denotes the derivative {\vf with respect to}  conformal Euclidean time $\eta_E$ and ${\cal H}_E^{\ } \equiv a'/a$.
In this regime the solutions that are regular at the SP take the approximate {\vf analytic} form
\be
f_{(n)} = \frac {\zeta_{(n)}(0)}{a} e^{n \eta_E}, \qquad a_{(n)} = -\frac{3\phi' \zeta_{(n)}(0)}{na} e^{n \eta_E},
\label{pertsol1}
\ee
where the constraint \eqref{pert3eq} was used to find the metric perturbation.
These solutions are valid in the complex $\eta$-plane in the regime $n \gg Ha$. 
From \eqref{zet1} it follows that in this regime, $\tilde \zeta_{(n)} \approx -Ha^3 f_{(n)}$.

One can verify whether the analytic approximations \eqref{pertsol1} are accurate by solving numerically for the perturbations simultaneously with the complex background. This can be done e.g. by integrating the field equations along a broken contour $C_B(X)$ in the complex $\tau$-plane that runs along the real axis to a point  $X$, and then up the imaginary $y$-axis. When $\phi_0 \ge \phi_0^c$ one can adjust both the turning point $X$ and the phase angle $\gamma$ of $\phi (0)$ so that $a$ and $\phi$ tend to real functions $b(y)$ and $\chi(y)$ along the vertical line given by $\tau = X + iy$ in the complex $\tau$-plane \cite{HHH08b}. {\vf These are the scale factor and scalar field of a classical Lorentzian solution. } An example of an exact complex background that tends to a classical history is shown in Figure \ref{background}, for $\phi_0=4$ and $m^2=.05$. 

\begin{figure}[t]
\includegraphics[width=3.2in]{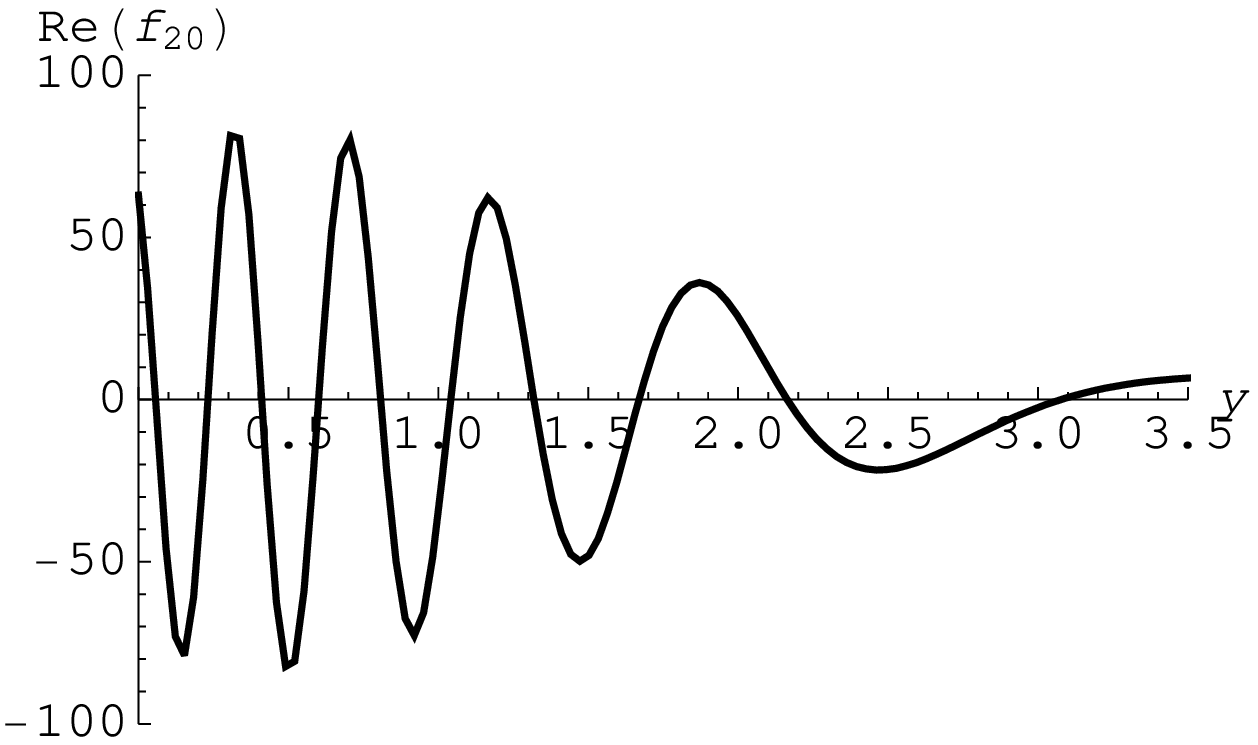}\hfill
\includegraphics[width=3.2in]{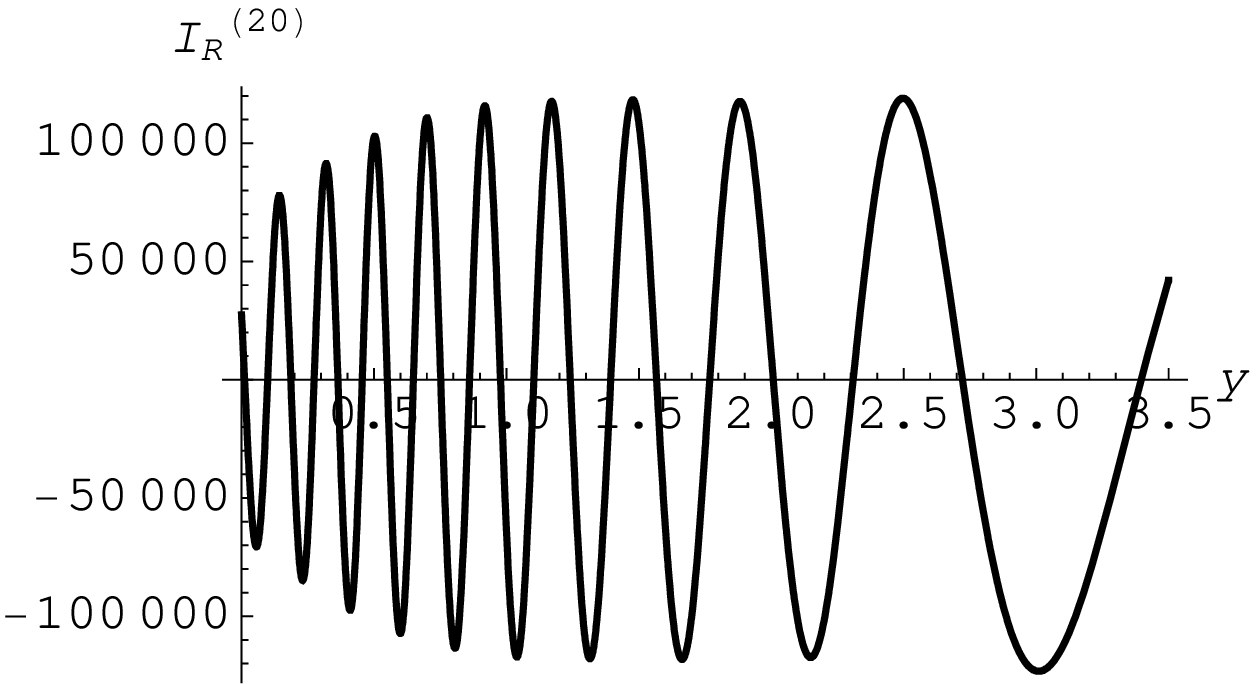} 
\caption{{\it Left panel}: When the perturbation mode is inside the horizon the complex scalar field fluctuation oscillates with amplitude $\propto a^{-1}$, as illustrated here for the real part of the $n=20$ mode in the background of Figure 1.\\ {\it Right panel}: As a consequence of this, the Euclidean action of a perturbation mode that is inside the horizon oscillates with an approximately constant amplitude.}
\label{pertmatter}
\end{figure}

In Figure \ref{pertmatter} (left panel) we plot the evolution of the $n=20$ matter perturbation along the vertical part of the contour in this background. The range of $y$ shown here corresponds to the regime where the mode is inside the horizon. One sees it oscillates rapidly with decreasing amplitude $\propto a^{-1}$, in good agreement with the analytic approximation \eqref{pertsol1}.
The Euclidean action of a solution to the equations \eqref{eucperteqns} is just a boundary term \cite{HLL93}, 
\be
I^{(n)} = M \tilde  z_{(n)} \tilde z_{(n)}' - N \tilde z_{(n)}^2
\label{pertact}
\ee
where
\be
M\equiv\frac{(n^2-4)}{2[(n^2-4)a'^2 + 3a^2 \phi'^2]}
\ee
and
\be
N\equiv \frac{1}{4MUa^3} \left[ K_n \left(2a^4 -3a^6 m^2 \phi^2 + 3 \frac{n^2-1}{n^2 -4} a^4 \phi'^2 \right) + a^{12} m^4 \phi^2 + 3 a^9 \phi \phi' a' \right]
\ee
with $U=K_n a a' + a^8 m^2 \phi \phi'$ and $K_n \equiv\frac{1}{3} \left[ (n^2 -4)a'^2 - (n^2 +5) a^4 \phi'^2 - (n^2 -4) a^6 m^2 \phi^2 \right]$. \\ 
All quantities here are evaluated on the boundary surface where one calculates the wave function. In the complex $\tau$-plane this surface is given by a certain value $\tau_f = X+iy_f$ where the variables take real values $a(\tau_f)=b$, $\phi(\tau_f)=\chi$ and $\tilde \zeta(\tau_f) = \tilde z$.

\begin{figure}[t]
\includegraphics[width=3.2in]{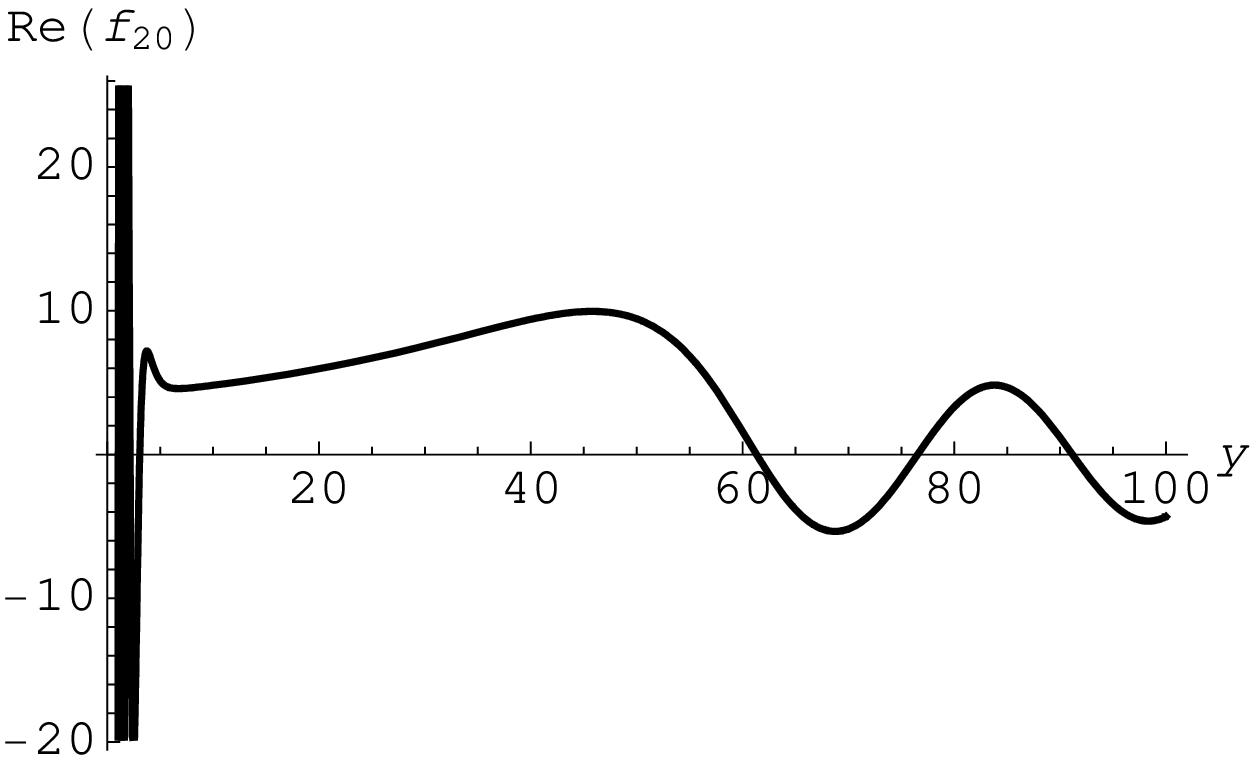} \hfill
\includegraphics[width=3.2in]{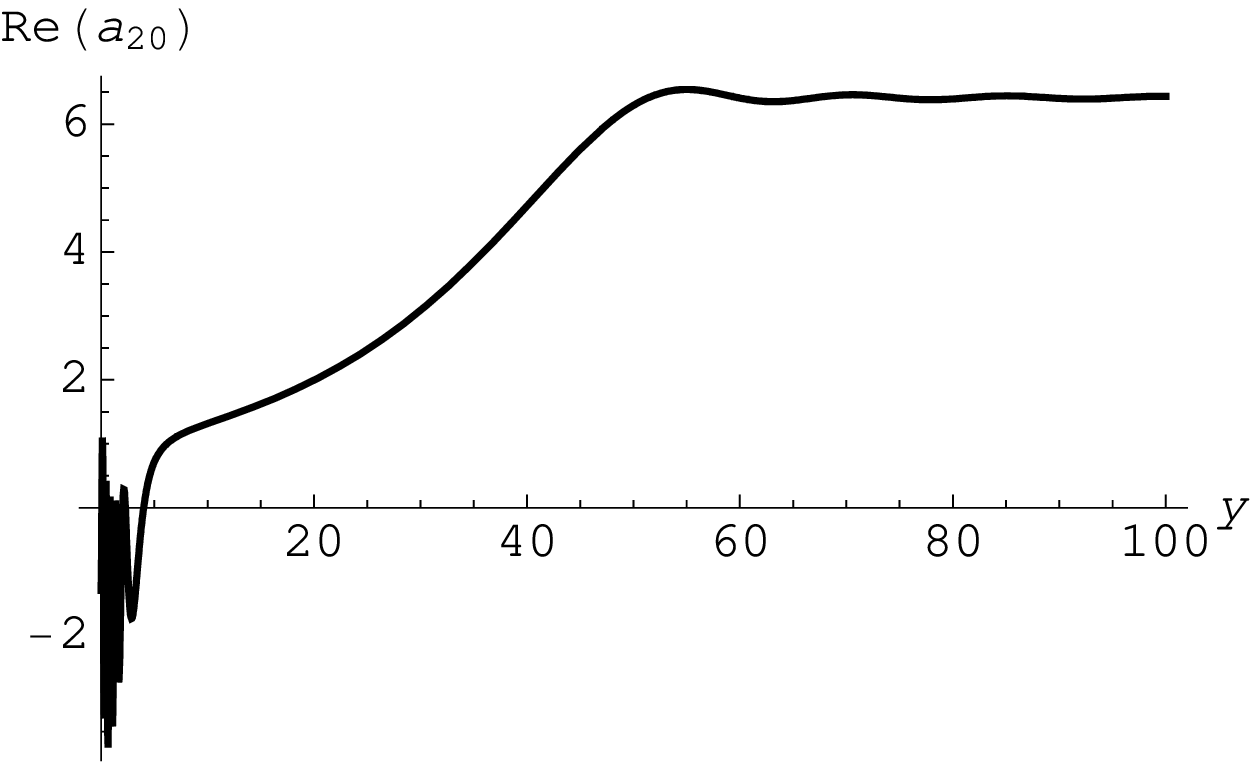} 
\caption{Numerical solution of the perturbation modes $a_{(n)}$ and $f_{(n)}$ for $n=20$ in the exact complex $\phi_0=4$ background shown in Figure \ref{background}. The modes are complex and oscillate when the absolute value of their wavelength is smaller than the Hubble radius, or $n> |aH_E^{\ }| $. When the wavelength crosses the Hubble radius around $y \sim 5$ both the matter - and metric perturbation start slowly growing until the end of inflation around $y \sim 60$. The imaginary part of the gauge invariant combination $\zeta_{(n)}$  decays away in this regime. After inflation ends the metric perturbation is essentially real and constant, with small oscillations due to the oscillating background scalar field. These primordial metric perturbations provide the seeds for structure formation in the corresponding Lorentzian cosmology.}
\label{pertmetric}
\end{figure}

When the absolute value of the wavelength $a/n$ of a complex perturbation mode becomes larger than the Hubble radius, both the scalar field and metric perturbations stop oscillating and start slowly growing. This transition can be clearly seen in the numerical solutions shown in Figure \ref{pertmetric}. It can also be understood analytically: Outside the horizon the gradient term is unimportant in the equations of motion \eqref{eucperteqns}, which therefore admit growing and decaying solutions for $a_{(n)}$ and $f_{(n)}$. The growing solutions are given by
\be
f_{(n)}^g \sim \frac{1}{\phi}, \qquad a_{(n)} = \frac{1}{\phi}f_{(n)}^g
\label{growsol}
\ee
and the decaying modes are
\be
f_{(n)}^d \sim \frac{1}{a^3}, \qquad a_{(n)} = -\frac{m}{2H}f_{(n)}^d.
\label{decsol}
\ee
The general solution for $\tilde \zeta_{(n)}$ in this regime is a combination of a growing and decaying mode.
The (complex) proportionality constants multiplying each term can be approximately determined in terms of $\zeta_{(n)}(0)$ by matching the solution on subhorizon scales at horizon crossing $n= a_{*} H_{*}$. Here the subscript star means the quantity is to be evaluated at the time of horizon crossing of modes with wavenumber $n$.
At horizon crossing $\tilde \zeta_{(n)}$ generally has an imaginary component, since the scalar field and metric perturbation are not simultaneously real. The requirement that $\tilde \zeta_{(n)}$ be real at the boundary essentially means that the phase $\theta$ of $\zeta_{(n)} (0)$ at the SP should be tuned so that the imaginary component of the subhorizon mode function matches onto the decaying mode when the perturbation leaves the horizon.  

\begin{figure}[t]
\includegraphics[width=3.2in]{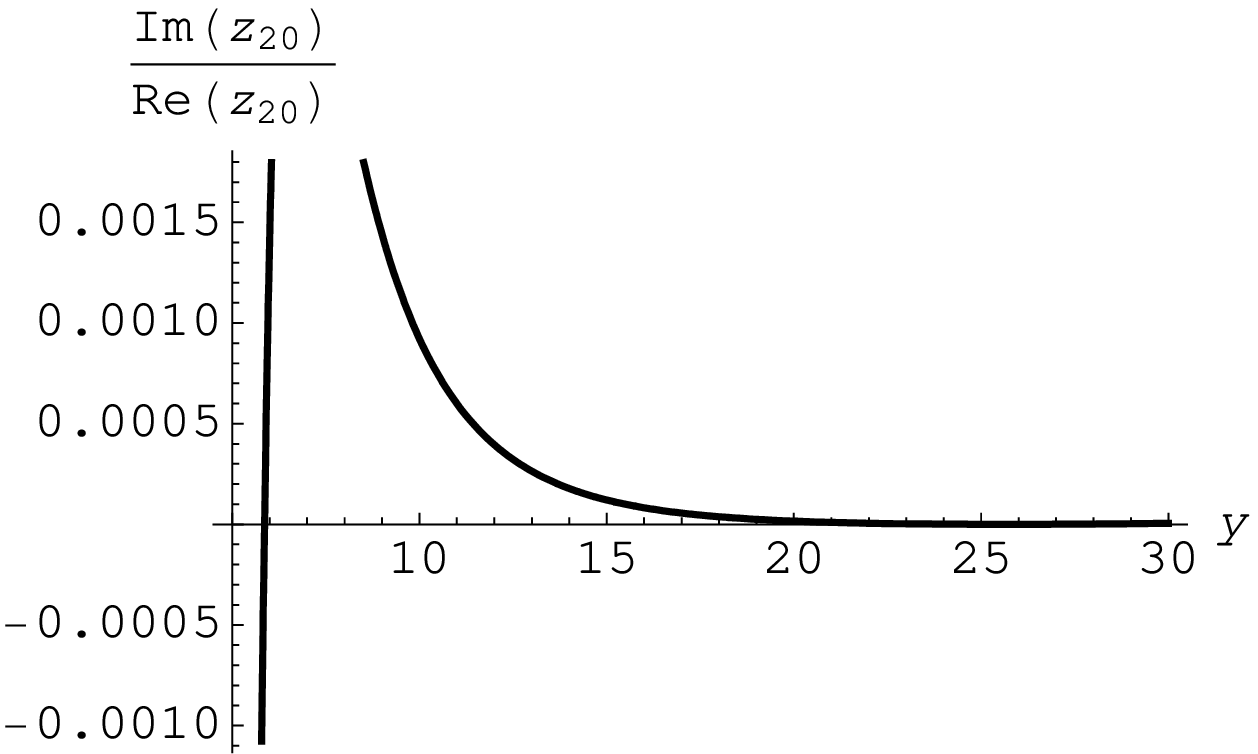} \hfill
\includegraphics[width=3.2in]{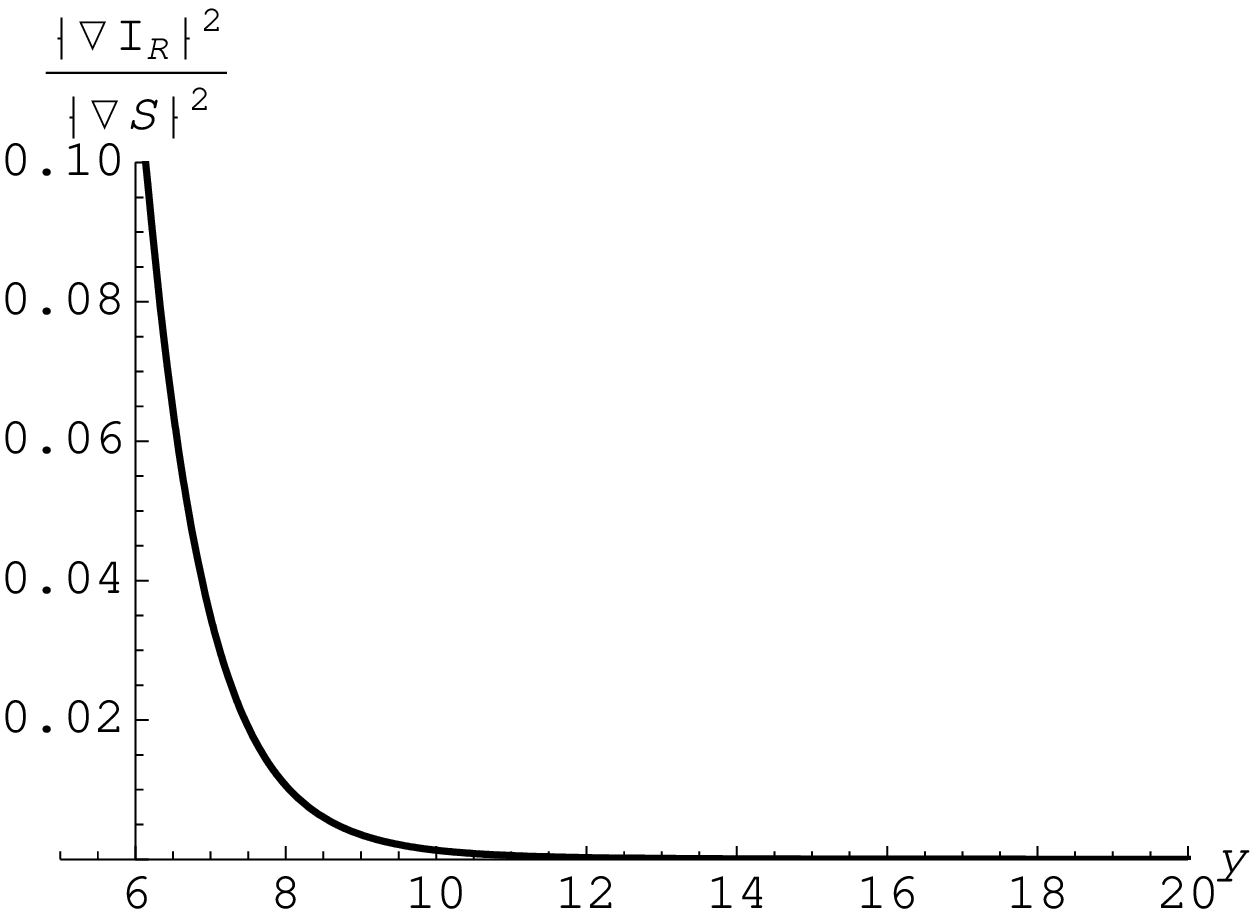} 
\caption{{\it Left panel}: The phase of the perturbation mode at the SP should be tuned so that $z_n$ is real at the boundary. This is illustrated here for the numerical perturbation solution $z_{20}^{\ }$ along the vertical part of a contour in the complex $\tau$-plane, in the complex $\phi_0=4$ background.\\  {\it Right panel:} The ratio of the gradients of the real part of the Euclidean action to the imaginary part tends to zero when the wavelength of a perturbation mode becomes larger than the Hubble radius.}
\label{classicality}
\end{figure}

This also means perturbations behave classically when their wavelength exceeds the Hubble radius, since the information on their phase decays away. We illustrate this in Figure \ref{classicality} where $\theta$ is fine-tuned so that the numerical solution $\tilde \zeta_{(n)}$, for $n=20$, tends to a real function $z_{(n)}$ along the vertical part of the broken contour $C_B(X)$ in the $\phi_0=4$ background and with $m^2 = .05$. One sees the ratio of the gradients of the real part of the Euclidean action to the imaginary part tends to zero. 

The real part of the action \eqref{pertact} tends to a constant\footnote{The  asymptotic value of $I_R$ can also be obtained from the approximate analytic form of the superhorizon solutions \eqref{growsol} and \eqref{decsol}. Indeed, while the growing term in $\tilde \zeta_{(n)}$ must be tuned to be real outside the horizon, the derivative $z_{(n)}' $ contains an imaginary component of order $a_{*}^2 \zeta_{(n)0}a H^2$ that arises from taking the derivative of the decaying mode. This is a subleading contribution to the total action, but it gives rise to a real part $I_R$ of the correct magnitude.}, which is approximately given by its value when the mode leaves the horizon. Hence for the approximate analytic solutions \eqref{pertsol1} we obtain
\be
I_R^{(n)} \rightarrow \frac{n\tilde z_{(n)}^2(y_{*})}{2b_{*}^4H_{*}^2} 
\label{Epertact}
\ee

\nopagebreak 

\end{document}